\documentclass[twocolumn]{bmcart}

\usepackage{amsmath,graphicx,cite}
\usepackage{amssymb,mathtools,bm,etoolbox,dblfloatfix,subcaption,siunitx}
\usepackage{booktabs}
\usepackage[referable]{threeparttablex} % for notes in tables
\usepackage{url}
\usepackage[
  justification=RaggedRight,
  calcwidth=.87\linewidth,
]{caption}

\begin{document}

\begin{frontmatter}

\begin{fmbox}
\dochead{Research}

\title{MYRiAD: \\ A Multi-Array Room Acoustic Database}

\author[
  addressref={aff1},                   % id's of addresses, e.g. {aff1,aff2}
  corref={aff1},                       % id of corresponding address, if any
% noteref={n1},                        % id's of article notes, if any
  email={thomas.dietzen@esat.kuleuven.be}   % email address
]{\inits{T.}\fnm{Thomas} \snm{Dietzen}}
\author[
  addressref={aff1},
  email={thomas.dietzen@esat.kuleuven.be}
]{\inits{R.}\fnm{Randall} \snm{Ali}}
\author[
  addressref={aff2},
%  email={john.RS.Smith@cambridge.co.uk}
]{\inits{M.}\fnm{Maja} \snm{Taseska}}
\author[
  addressref={aff1},
  email={toon.vanwaterschoot@esat.kuleuven.be}
]{\inits{T.}\fnm{Toon} \snm{van Waterschoot}}

\address[id=aff1]{%                           % unique id
  \orgdiv{Dept. of Electrical Engineering (ESAT), STADIUS Center for Dynamical Systems, Signal Processing and Data Analytics},             % department, if any
  \orgname{KU Leuven},          % university, etc
  \city{Leuven},                              % city
  \cny{Belgium}                                    % country
}
\address[id=aff2]{%
 % \orgdiv{},
  \orgname{Microsoft},
  %\street{},
  %\postcode{}
  \city{Munich},
  \cny{Germany}
}

\begin{abstractbox}

\begin{abstract} 
In the development of acoustic signal processing algorithms, their evaluation in various acoustic environments is of utmost importance. In order to advance evaluation in realistic and reproducible scenarios, several high-quality acoustic databases have been developed over the years. In this paper, we present another complementary database of acoustic recordings, referred to as the Multi-arraY Room Acoustic Database (MYRiAD). The MYRiAD database is unique in its diversity of microphone configurations suiting a wide range of enhancement and reproduction applications (such as assistive hearing, teleconferencing, or sound zoning), the acoustics of the two recording spaces, and the variety of contained signals including 1214 room impulse responses (RIRs), reproduced speech, music, and stationary noise, as well as recordings of live cocktail parties held in both rooms. The microphone configurations comprise a dummy head (DH) with in-ear omnidirectional microphones, two behind-the-ear (BTE) pieces equipped with 2 omnidirectional microphones each, 5 external omnidirectional microphones (XMs), and two concentric circular microphone arrays (CMAs) consisting of 12 omnidirectional microphones in total. The two recording spaces, namely the SONORA Audio Laboratory (SAL) and the Alamire Interactive Laboratory (AIL), have reverberation times of \SI{2.1}{s} and \SI{0.5}{s}, respectively. Audio signals were reproduced using 10 movable loudspeakers in the SAL and a built-in array of 24 loudspeakers in the AIL. MATLAB and Python scripts are included for accessing the signals as well as microphone and loudspeaker coordinates. The database is publicly available at \cite{Dietzen2022zenodo}.

\end{abstract}

\begin{keyword}
\kwd{Room Acoustic Database}
\kwd{Room Impulse Response}
\kwd{Cocktail Party Noise}
\kwd{Microphone Array}
\kwd{Loudspeaker Array}
\kwd{Acoustic Signal Processing}
\end{keyword}

\end{abstractbox}

\end{fmbox}
\end{frontmatter}

\section{Introduction}
\label{sec:intro}

Acoustic signal processing using multiple microphones has received significant attention due to its fundamental role in a number of applications such as assistive hearing with hearing aids or cochlear implants, teleconferencing, hands-free telephony, voice-controlled devices, spatial audio reproduction, and sound-zoning, just to name a few. 
Some of the specific tasks which can be accomplished with acoustic signal processing include speech enhancement and speech dereverberation \cite{loizou2007speech, gannot07, doclo2010acoustic, naylor10, brandstein2013microphone, kinoshita2016summary, gannot2017consolidated, SourceSep2018},  room parameter estimation \cite{eaton2014_ACE}, acoustic echo and feedback cancellation \cite{sridhar2021icassp, vanWaterschoot2010fifty}, source localisation \cite{gannot07, brandstein2013microphone, evers2020locata}, audio source separation \cite{gannot2017consolidated, SourceSep2018}, sound field control \cite{coleman2014acoustic, betlehem2015personal}, and automatic speech recognition \cite{chime5_18}, all of which are pertinent to the aforementioned applications.
One of the core phases in the development of acoustic signal processing algorithms is that of the evaluation phase, where the performance of a newly developed algorithm is compared to that of existing algorithms in various acoustic environments which are relevant for the application at hand. This is clearly challenging because the laboratory conditions under which the algorithm is evaluated rarely match the real-world conditions where the algorithm must perform. Additionally, recorded audio signals with the target microphone configurations and specified acoustic scenarios may be unavailable, resulting in the use of simulated data for evaluation. Although simulated data can be useful in the evaluation of initial proof of concept ideas, it does not necessarily provide accurate indication whether the algorithm will perform well in real-world conditions.
In an effort to overcome these challenges and to encourage the use of more realistic data, several high-quality acoustic databases 
containing room impulse responses (RIRs) \cite{wen2006evaluation, Jeub2009, Kayser2009, stewart2010database, Nielsen2014, Hadad2014, eaton2014_ACE, Woods2015, kinoshita2016summary, szoke2019building, carlo2021dechorate, vcmejla2021mirage, Koyama2021, Zhao2022}, speech \cite{Nielsen2014, eaton2014_ACE, Woods2015, kinoshita2016summary, chime5_18, szoke2019building, sridhar2021icassp}, music \cite{Nielsen2014}, and babble or cocktail party noise \cite{Woods2015, van2019dipco, Fischer2020} have been developed over the years, which have played an important role in building confidence in the real-world performance of various acoustic signal processing algorithms. 

In this paper, we present another complementary database of acoustic recordings from multiple microphones in various acoustic scenarios, referred to as the Multi-arraY Room Acoustic Database (MYRiAD). 
In comparison to the existing databases, the MYRiAD database is unique in its diversity of the employed microphone configurations suiting a wide range of applications, the acoustics of the  recording spaces, and the variety of signals contained in the database, which includes RIRs, recordings of reproduced speech, music, and stationary noise, as well as recordings of live cocktail parties.

The database consists specifically of two different microphone configurations used across two different rooms. 
The first microphone configuration consists of a dummy head (DH) with in-ear omnidirectional microphones, two behind-the-ear (BTE) pieces mounted on the DH, each equipped with 2 omnidirectional microphones,\footnote{BTE pieces are commonly used for hearing aids or cochlear implant devices. There is no additional processing done on the microphone signals before arriving to the data acquisition system.} as well as 5 external omnidirectional microphones (XMs) located at various distances and angles from the DH.{\footnote{A typical use-case of such XMs consists in providing additional information to improve the enhancement of BTE signals \cite{Farmani2017, goessling2020, ali2020multi}.}}
This microphone configuration will be referred to as M1.
The second microphone configuration consists of two concentric circular microphone arrays (CMAs) with in total 12 omnidirectional microphones,{\footnote{ 
Among others, use-cases of CMAs include signal enhancement \cite{Huang2017} and localization \cite{Pavlidi2013}, for instance in smart speakers, and sound zoning \cite{coleman2014acoustic, betlehem2015personal}.
}} which will be referred to as M2. 
The two different rooms where audio recordings  were  made are: (i) the SONORA Audio Laboratory \cite{vanWaterschoot2022labs} located at the Depeartment of Electrical Engineering (ESAT-STADIUS), KU Leuven, Belgium, which we will refer to as the SAL, and (ii) the Alamire Interactive Laboratory \cite{vanWaterschoot2022labs} located at the Park Abbey in Heverlee, Belgium, referred to as the AIL. The main acoustical difference between these two rooms is that the SAL is significantly more reverberant than the AIL, with reverberation times of \SI{2.1}{s} and \SI{0.5}{s}, respectively. 
In the SAL, the microphone configuration M1   was   used in one position, and in the AIL, a combination of microphone configurations M1 and M2   was   used in two positions. 
In terms of sound generation, 10 different movable loudspeakers   were   used as artificial sound sources in the SAL, while the AIL has been equipped with an array of 24 loudspeakers. 

\begin{figure}
	\centering
	\includegraphics[scale = 0.36]{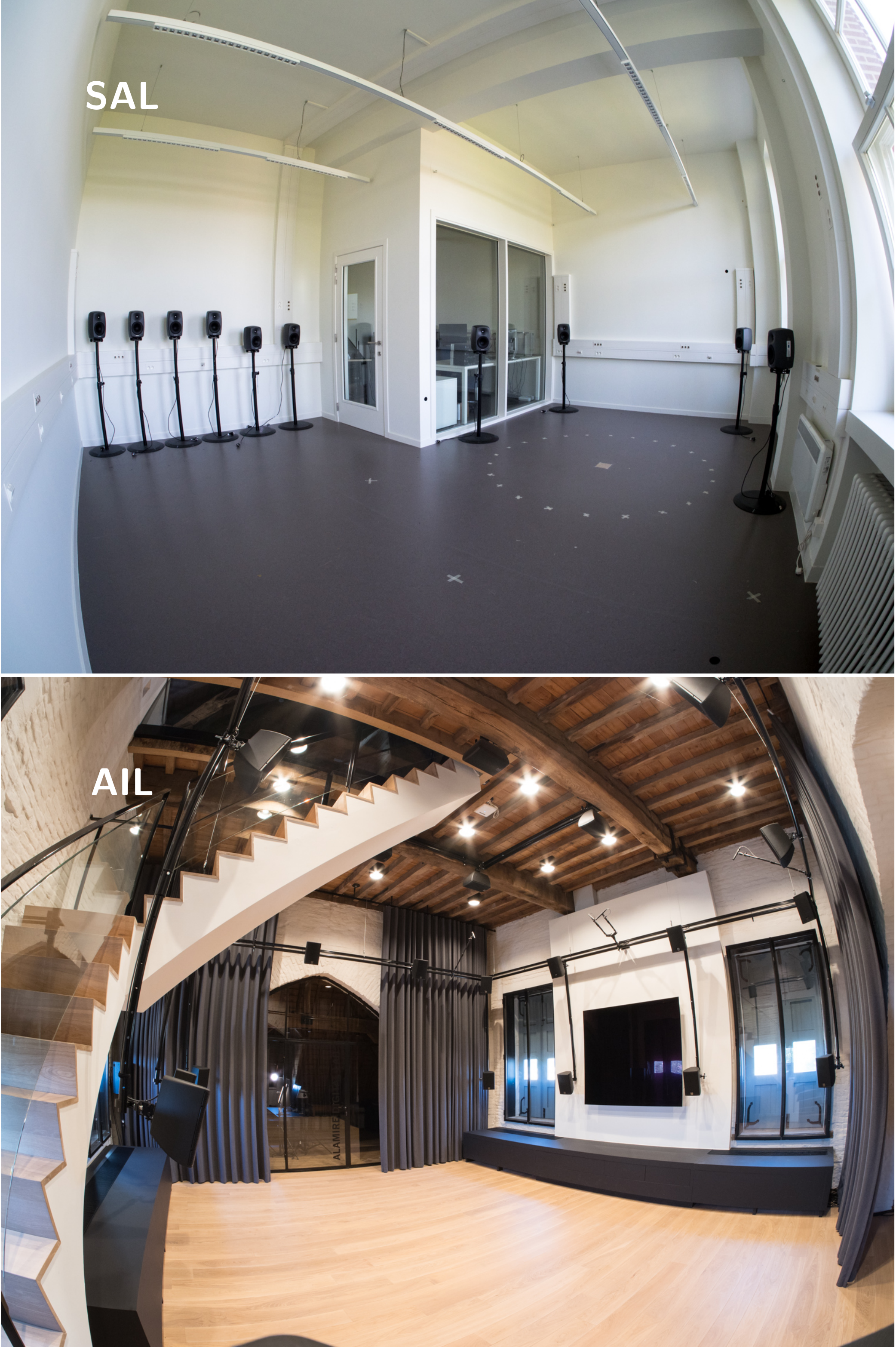}
	\captionsetup{oneside,margin={0.16cm, 0.5cm}}	
	\caption{\small Fisheye view of the SAL and the AIL.}
	\label{fig:fisheye_both}
\end{figure}

\renewcommand{\arraystretch}{1.4}
\begin{table*}[t] 
	\caption{\small Equipment used for creating the database.}
	\begin{center}
		\begin{threeparttable}	
			\begin{tabular}{@{}l l l l l} 
				\toprule
				Type & & & Product & Room/\\
				& & & & Mic. Config.\\
				\midrule
				Hardware & Reproduction & Loudspeakers & Genelec 8030 CP & SAL \\ 
				& & & Martin Audio CDD6 & AIL \\
				\cmidrule{3-5}
				& & DA-converters & RME M-32 DA & SAL\\
				& & & Powersoft OTTOCANALI 4K4 DSP+D & AIL\\
				\cmidrule{2-5}
				& Acquisition & Microphones &  Neumann KU-100 DH &M1\\
				& & &  BTE left/right-ear pieces from Cochlear  &  M1\\
				& & &  AKG {CK97-O}  &M1\\
				& & & AKG CK32  &M1 \& M2\\
				& & & DPA 4060 & M2\\
				\cmidrule{3-5}
				& & AD-converters/pre-amplfiers & RME Micstasy & SAL \& AIL\\
				& & & Proprietary pre-amp. for BTE microphones & SAL \& AIL \\
				\cmidrule{2-5}
				& Digital interface & & RME Digiface USB audio interface & SAL \\ 
				& & & Ferrofish Verto 64 & AIL \\ 
				& & & Apple iMac & SAL \& AIL \\
				\midrule
				Software &  Reproduction/acquisition & &  Logic Pro X & SAL \\
				& & & Adobe Audition & AIL \\
				\cmidrule{2-5}
				& Post-processing & & MATLAB \\
				& & & Python \\
				\hline
			\end{tabular}
			
		\end{threeparttable}
		
	\end{center}
	\label{tab:equip}
\end{table*}

The following audio signals   were   played back through the speakers and recorded by the microphones: exponential sine sweeps used to compute RIRs \cite{Holters2009} between source and microphone positions, resulting in 110 RIRs for the SAL and 1104 RIRs for the AIL, as well as three male speeches  \cite{VCTK}, three female speeches  \cite{VCTK}, a drum beat  \cite{AECW2011}, a piano piece  \cite{EBU}, and speech-shaped stationary noise.
Additionally, in both rooms, several participants   were   invited to re-create a live cocktail party scenario. The resulting noise from the different cocktail parties held at each of the spaces   was   recorded for both microphone configurations. 

In total, the MYRiAD database contains 76 hours of audio data sampled at \SI{44.1}{\kHz} in \SI{24}{bit}, which results in \SI{36.2}{GB}.
All computed RIRs and recorded signals are available in the database and can be downloaded \cite{Dietzen2022zenodo}. MATLAB and Python scripts are included in the database for accessing the signals and corresponding microphone and loudspeaker coordinates.

The remaining sections of this paper provide a detailed overview of the database and are organised as follows.
 In Sec. \ref{sec:room}, an overview of the two different rooms, the SAL and the AIL, is presented. 
 In Sec. \ref{sec:equipment}, a detailed description is given of the equipment used.
 In Sec. \ref{sec:MicLSsetup}, the microphone and loudspeaker configurations within the two rooms are discussed.
 In Sec. \ref{sec:recorded}, an overview is given of the recorded signals, details of the cocktail party, and the computed RIRs.
 In Sec. \ref{sec:examples}, practical instructions for using the database are provided, along with a description of relevant MATLAB and Python scripts, and some examples from the database are illustrated. 
 In Sec. \ref{sec:conc}, the database is briefly summarised.

 \begin{figure}
	\centering
	\includegraphics[scale = 0.29]{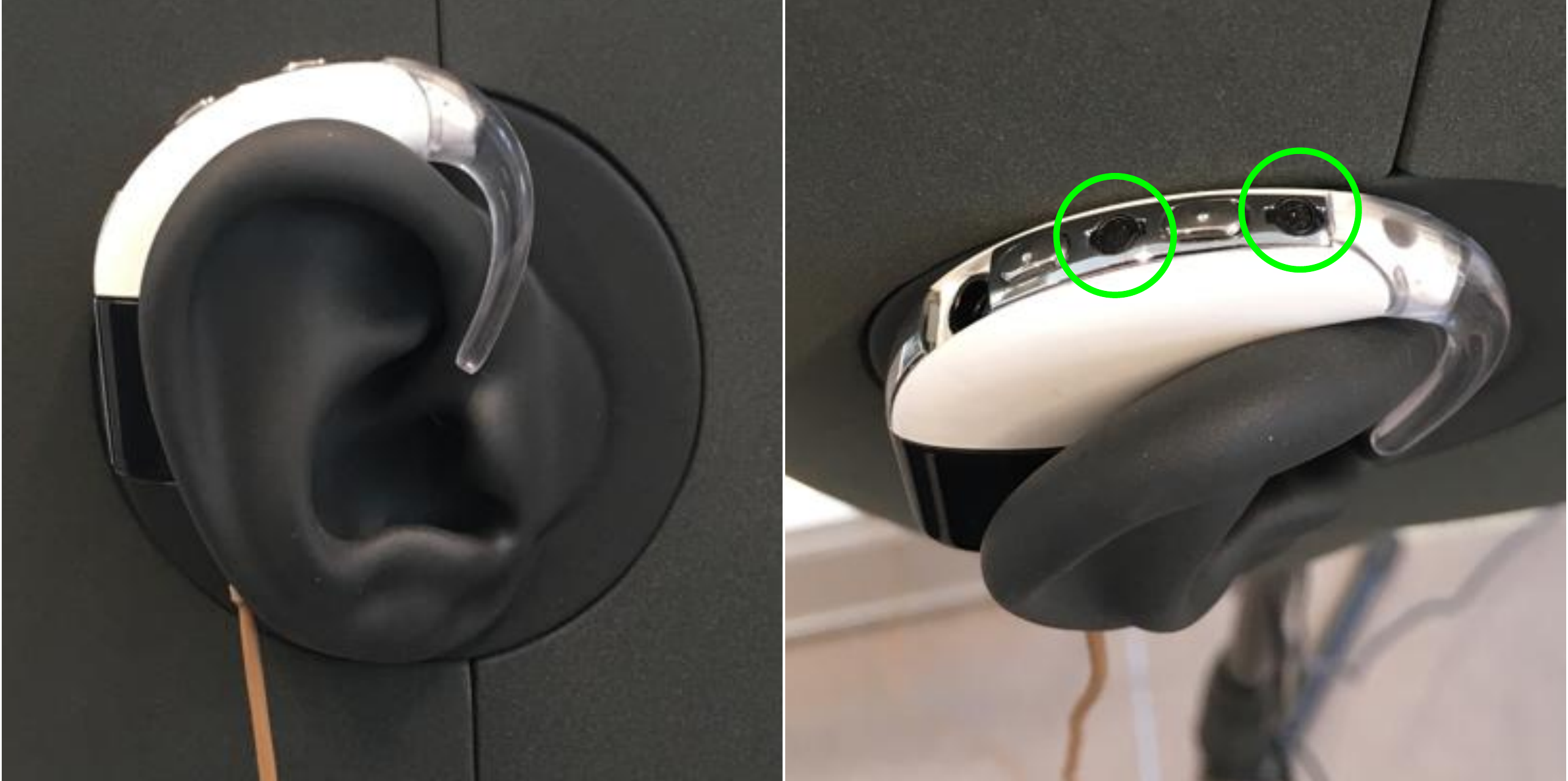}
	\captionsetup{oneside,margin={0.16cm, 0.5cm}}	
	\caption{\small Dummy BTE pieces used for creating the database. Each BTE piece consists of two omnidirectional microphones as indicated by the circles.}
	\label{fig:bteha}
\end{figure}

\begin{figure*}[t]
	\includegraphics[scale = 0.175]{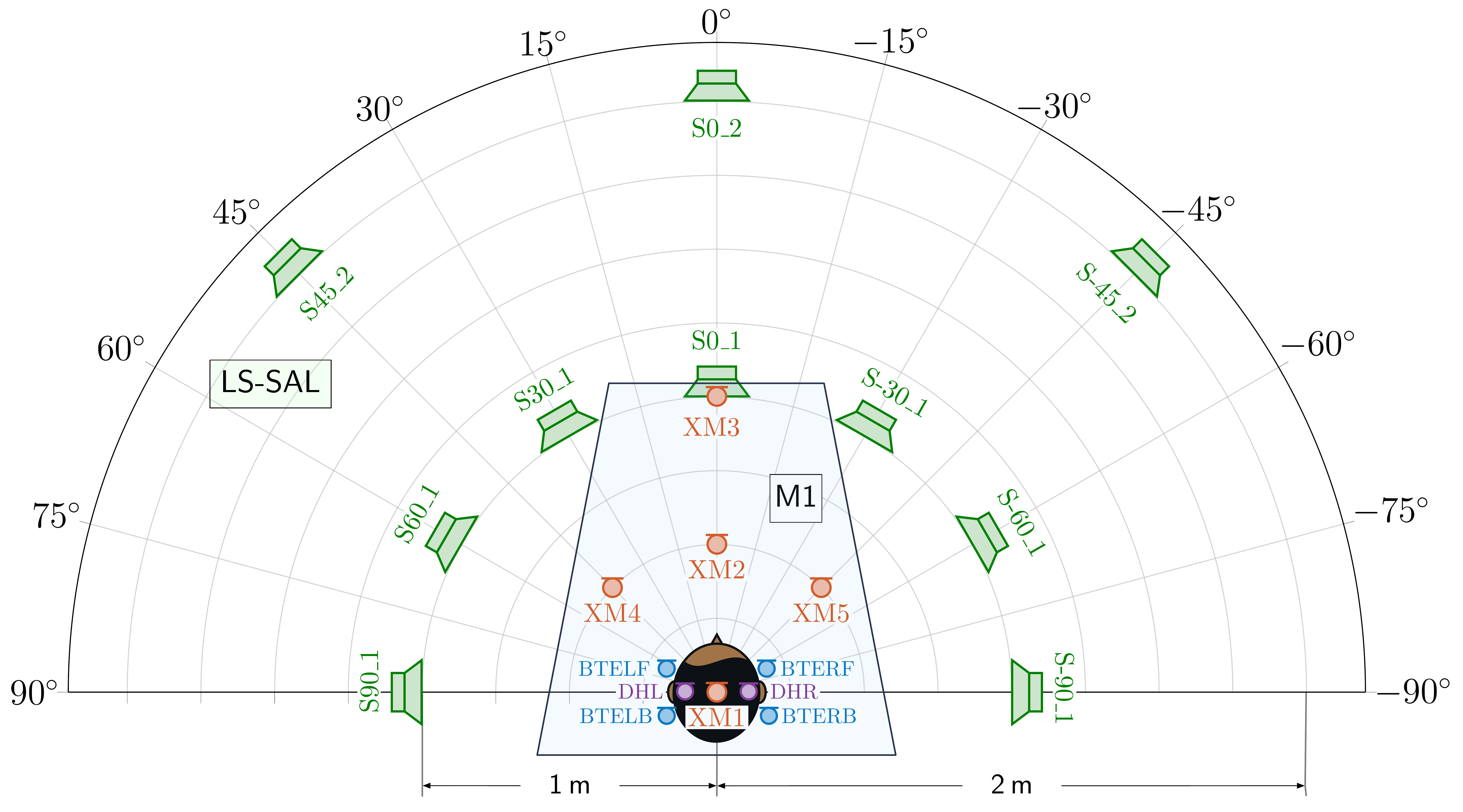}
	\captionsetup{oneside,margin={.16cm, .2cm}}	
	\begin{minipage}[t]{2\columnwidth}
		\caption{\small Plan view of the M1  microphone configuration and the LS-SAL loudspeaker configuration. A description of the microphone and loudspeaker labels is given in Table \ref{tab:mic_ls_labels}. The radial grid spacing of the polar plot is \SI{0.25}{\metre}. The DH is placed at a height of approximately \SI{1.3}{\meter} ear level from the floor and all XMs are placed at a height approximately \SI{1}{\meter} from the floor. The trapezoidal shape is used to represent the M1 microphone configuration in the floor plans of Fig. \ref{fig:mic_ls_placement}. For extracting the coordinates of the microphone and loudspeaker positions, the MATLAB or Python scripts discussed in Sec. \ref{sec:retrievingcoordinates} should be used.}
				\label{fig:MicSetupM1}
	\end{minipage}
\end{figure*}

\section{Room description}
\label{sec:room}

In this section, we provide a brief overview on the characteristics of the two recording rooms.
The SAL is described in Sec. \ref{sec:room_description_SAL} and the AIL is described in Sec. \ref{sec:room_description_AIL}.

\subsection{SONORA Audio Laboratory (SAL)}
\label{sec:room_description_SAL}

The SAL \cite{vanWaterschoot2022labs} is located at the Department of Electrical Engineering (ESAT-STADIUS), KU Leuven, Heverlee, Belgium.
Fig. \ref{fig:fisheye_both} shows a fisheye view and Fig. \ref{fig:mic_ls_placement} shows a floor plan of the L-shaped SAL with approximate dimensions. The height of the room is  $\SI{3.75}{\metre}$, yielding a volume of approximately $\SI{102}{\metre^3}$.
The walls and ceiling are made of plasterboard covering mineral wool, while the floor is made of concrete covered with vinyl.
Two windows, each of \SI{4}{\metre^2} are located on one side of the room.
Adjacent to the recording room, separated by glass of area \SI{6.5}{\metre^2}, is the control room, where all the acquisition equipment and a computer are located. From the RIRs measured in the SAL, we estimated the reverberation time  {$\mathrm{T_{20}}$} to be \SI{2.1}{\s} as described in Sec. \ref{sec:reverbtimes}.
Details on the audio hardware used in the SAL are given in Sec. \ref{sec:equipment}, while the microphone and loudspeaker configuration and placement are described in  Sec. \ref{sec:M1}, Sec. \ref{sec:LS_SAL}, and Sec. \ref{sec:setupplacement}.

\subsection{Alamire Interactive Laboratory (AIL)}
\label{sec:room_description_AIL}

The AIL \cite{vanWaterschoot2022labs} is located in a historic gate building, the Saint Norbert's gate of the Park Abbey in Heverlee, Belgium.
Fig. \ref{fig:fisheye_both} shows a fisheye view and Fig. \ref{fig:mic_ls_placement} shows a floor plan of the room.
Apart from a staircase leading to a floor above, the room is approximately shoe-box shaped with \SI{6.4}{\metre} width, \SI{6.9}{\metre} depth, and \SI{4.7}{\metre} height, yielding a volume of approximately \SI{208}{\metre^3}.
The floor and ceiling are made of wood. 
The room is closed by thin line plastered brick walls with two windows each to the front and the back of about  \SI{3.3}{\metre^2} each, and wide passages to adjacent rooms, with one of them closed by a glass door.
These passages were closed off with curtains during recording, except for a part of the cocktail party noise, cf. Sec. \ref{sec:cocktail_party}.
The housing of the staircase is plastered, the stairs are wooden, and the railing is made of glass.
From the RIRs measured in the AIL, the reverberation time {$\mathrm{T_{20}}$} is estimated to be \SI{0.5}{\s}, cf. Sec \ref{sec:reverbtimes}. 
The AIL is equipped with a permanent, fixed array of 24 loudspeakers for spatial audio reproduction as shown in Fig. \ref{fig:fisheye_both}.
Further details on the audio hardware used in the AIL are given in Sec. \ref{sec:equipment}, while the microphone and  loudspeaker configuration and placement are described in  Sec. \ref{sec:M1},   Sec. \ref{sec:M2}, Sec. \ref{sec:LS_AIL}, and Sec. \ref{sec:setupplacement}.

\section{Recording equipment}
\label{sec:equipment}

A list of the recording and processing equipment used to create the database is shown in Table \ref{tab:equip}. In regards to the microphones, the DH contains 2 in-ear omnidirectional microphones (one for each ear) and the two BTE pieces (one for each ear) are each equipped with 2 omnidirectional microphones. The BTE pieces and their proprietary pre-amplifier were provided by Cochlear Ltd. and shown in Fig. \ref{fig:bteha}.
The specific loudspeaker and microphone configurations used for the various recordings in the database will be outlined in Sec. \ref{sec:MicLSsetup}, and naming conventions of files will be defined in Sec. \ref{sec:examples}.

The recording chains   were   built as follows.
As the digital audio workstations for sending and acquiring the signals, Logic Pro X and Adobe Audition on an iMac   were   used in the SAL and the AIL, respectively. 
In the SAL, the signals were sent from Logic Pro X via USB to the RME Digiface, then to the RME M-32 DA using the ADAT protocol, and finally to the respective Genelec 8030 CP loudspeakers. 
In the AIL, the signals were sent  from Adobe Audition via the DANTE protocol to the Powersoft OTTOCANALI 4K4 DSP+D, and finally to the Martin Audio CDD6 loudspeakers. 
In both rooms, all microphone signals were sent to an RME Micstasy (except for the BTE microphone signals which were firstly routed to the proprietary pre-amplifier) and converted to ADAT.
In the SAL, the ADAT signals were sent to the RME Digiface and finally recorded on Logic Pro X, whereas in the AIL, the ADAT signals were sent to the Ferrofish Verto 64 and via DANTE to Adobe Audition.
The various types of recorded signals are outlined in Sec. \ref{sec:recorded}.
For post-processing (such as RIR computation, cf. Sec. \ref{sec:recorded}), MATLAB and Python were used.

\section{Microphone and loudspeaker configurations}
\label{sec:MicLSsetup}

This section describes the microphone configurations in Sec. \ref{sec:microphone_setups}, the loudspeaker configurations in Sec. \ref{sec:loudspeaker_setups}, and the placement of these configurations within the SAL and AIL in Sec. \ref{sec:setupplacement}. The exact coordinates of the loudspeaker and microphone positions within the SAL and AIL from the various configurations can be loaded from the database, but the details of this procedure will be elaborated upon in Sec. \ref{sec:examples}.

\subsection{Microphone configurations}
\label{sec:microphone_setups}

\subsubsection{M1}
\label{sec:M1}

The first microphone configuration, M1, consists of the in-ear microphones from the DH, the microphones from the BTE pieces, three AKG {CK97-O} microphones, and two AKG CK32 microphones. As the AKG {CK97-O} and AKG CK32 microphones are not mounted on the DH, they are considered to be `external' in relation to the DH, and hence will be referred to as external microphones (XMs). 
This M1 configuration   was   used in both the SAL and the AIL, cf. Sec. \ref{sec:setupplacement}. 
Fig. \ref{fig:MicSetupM1} depicts the plan view of the measurement configuration of the loudspeakers and microphones used for the audio recordings made in the SAL. For now, however, we will focus only on the trapezoidal shape enclosing the microphones, which is a depiction of the M1 configuration. A description of the corresponding microphone labels is given in Table \ref{tab:mic_ls_labels}.

For this M1 configuration, the DH is placed at a height of approximately \SI{1.3}{\meter} ear level from the floor.
Each of the BTE pieces is mounted on the DH as shown in Fig. \ref{fig:bteha}.
The XMs are placed{\footnote{Note that XM1 is taped on a stand of \SI{18}{mm} diameter (holding the DH), which may impact the effective directivity pattern of the microphone at high frequencies.}} within a radius of \SI{1}{\metre} from the DH as shown in Fig. \ref{fig:MicSetupM1}. XM1, XM2, and XM3 are AKG {CK97-O} microphones, while XM4 and XM5 are AKG CK32 microphones. The XMs are all positioned at \SI{1}{\metre} above the floor.

\begin{table*}[t]
	\caption{Microphone and loudspeaker labels.}
	\begin{center}
		\label{tab:mic_ls_labels}
		\begin{tabular}{@{} l l l l } 
			\toprule
			& type & label & description \\ 
			\cmidrule{2-4}
			microphones & dummy head & DHL & left ear\\
			& & DHR & right ear\\
			\cmidrule{2-4}
			& BTE pieces & BTELF & left ear, front \\
			& & BTELB & left ear, back\\
			&& BTERF & right ear, front\\
			& & BTERB & right ear, back\\
			\cmidrule{2-4}
			& external microphone & XM[i] & with index [i] as depicted in Fig. \ref{fig:MicSetupM1}\\
			& & & [i] $\in$ \{1, 2, \dots, 5\}\\
			\cmidrule{2-4}
			& circular microphone array & CMA[r]\_[a] & at [r]$\si{cm}$ radius and an angle of [a]$^\circ$ as depicted in Fig. \ref{fig:MicSetupM2}\\
			& & & [r] $\in$ \{10, 20\}\\
			& & & [a] $\in$ \{-135, -90, \dots, 180\}\\
			\midrule
			\midrule
			& room & label & description \\ 
			\cmidrule{2-4}
			loudspeakers & SAL & S{[a]}\_{[d]} & at angle of {[a]}$^\circ$ and in {[d]}\SI{}{\meter} distance as depicted in Fig. \ref{fig:MicSetupM1}\\
				&  &  & {[a]} $\in$ \{-90, -60, -45, -30, 0, 30, 45, 60, 90\}\\	
				&  &  & {[d]} $\in$ \{1, 2\}\\		
			\cmidrule{2-4}
		   & AIL & S{[l]}{[i]} & at height level {[l]} with index {[i]} as depicted in Fig. \ref{fig:speakersAIL}\\
		   &  &  & {[l]} $\in$ \{L, U, T\} (indicating lower, upper, and top level)\\	
		   & &  & {[i]} $\in$ \{1, 2, \dots, 12\}\\	
			\bottomrule
		\end{tabular}
	\end{center}
\end{table*}

\begin{figure}
	\centering
	\includegraphics[scale = 0.175]{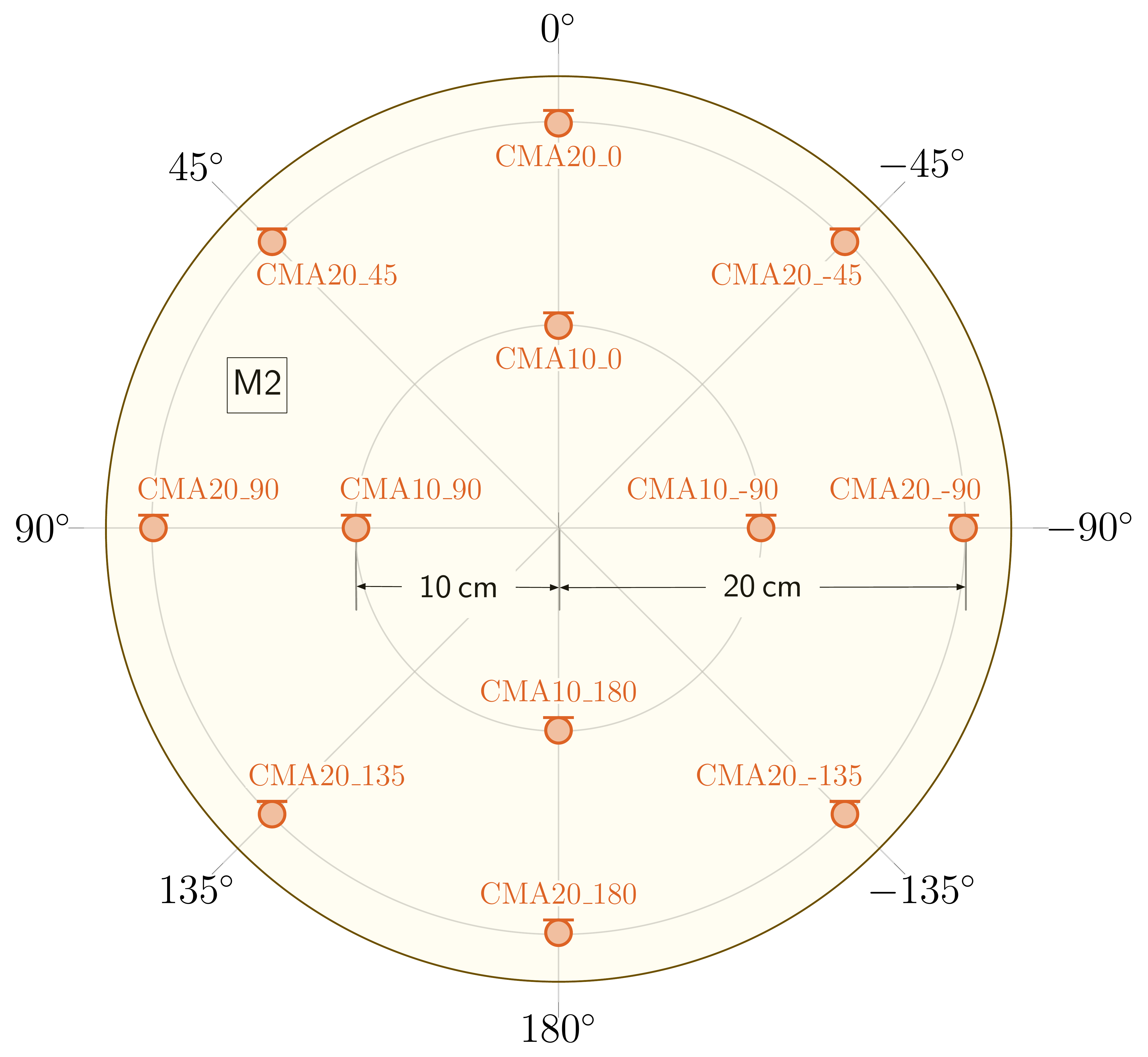}
	\captionsetup{oneside,margin={0.16cm, 0.5cm}}	
	\caption{\small Plan view of the M2 microphone configuration. A description of the microphone labels is given in Table \ref{tab:mic_ls_labels}. The radial grid spacing of the polar plot is \SI{0.1}{\metre}. DPA 4060 microphones are used for the inner circular microphone array and AKG CK 32 microphones are used for the outer circular microphone array. The circle drawn around the microphones represents the M2 microphone configuration in the floor plans in Fig. \ref{fig:mic_ls_placement}. For extracting more precise coordinates of the microphone and loudspeaker positions, the MATLAB or Python scripts discussed in Sec. \ref{sec:retrievingcoordinates} should be used.}
	\label{fig:MicSetupM2}
\end{figure}

\begin{figure*}[t]
	\includegraphics[scale = 0.175]{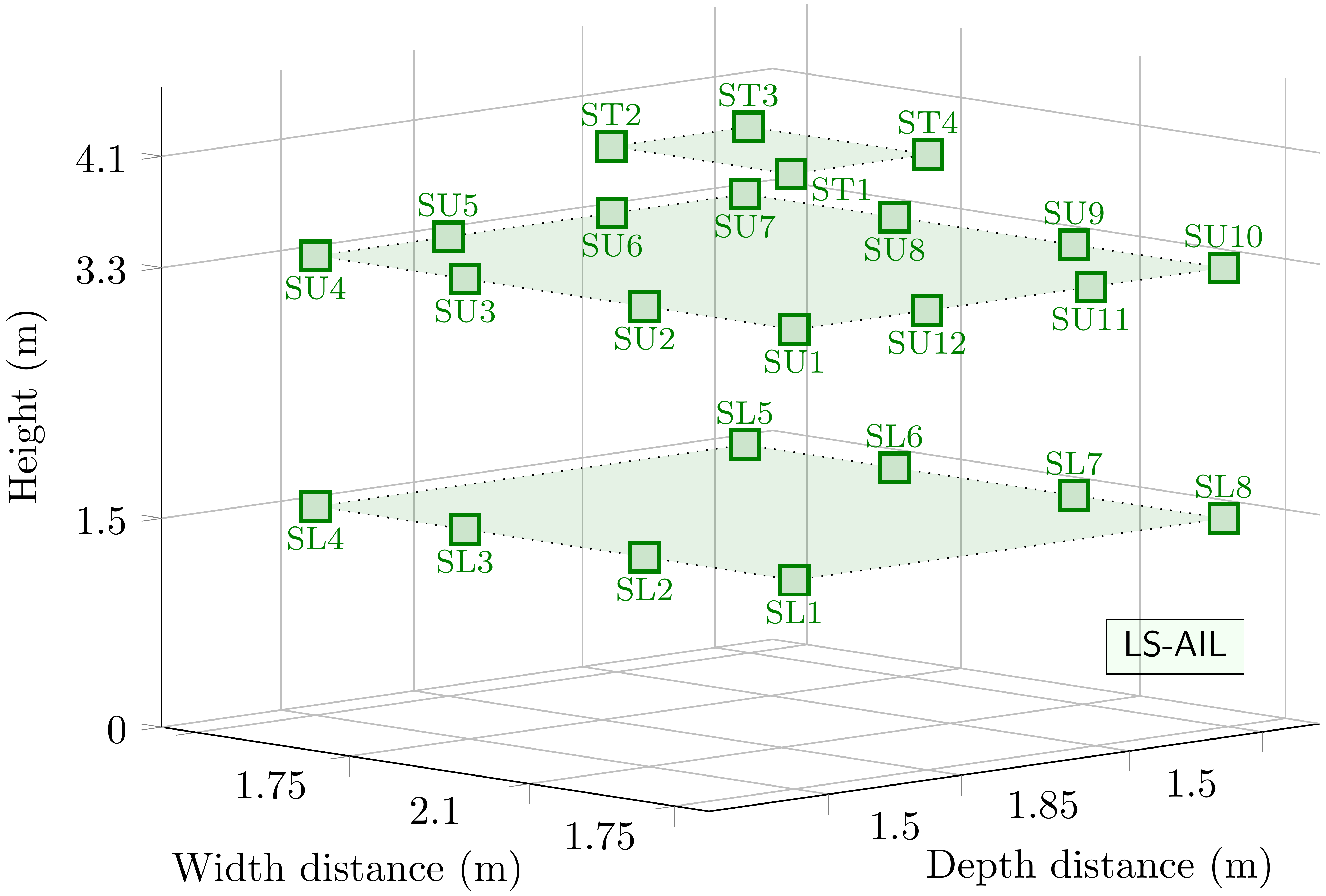}
	\captionsetup{oneside,margin={.16cm, .2cm}}	
	\begin{minipage}[t]{2\columnwidth}
		\caption{\small View of the  LS-AIL loudspeaker array in the AIL. 
		A description of the loudspeaker labels is given in Table \ref{tab:mic_ls_labels}.
		The speakers are organized in three different height levels of about \SI{1.5}{\metre} (lower level), \SI{3.3}{\metre} (upper level), and \SI{4.1}{\metre} (top level) above the floor.
		The axes limits coincide with the boundaries of the approximately shoe-boxed shaped room, cf. Sec. \ref{sec:room_description_AIL}. 
		On the horizontal axes, the approximate distance between neighbouring speakers is indicated.
		The given dimensions are of indicative nature and not exact; for extracting the coordinates of the microphone and loudspeaker positions, the MATLAB or Python scripts discussed in Sec. \ref{sec:retrievingcoordinates} should be used.}
			\label{fig:speakersAIL}
	\end{minipage}
\end{figure*}

\subsubsection{M2}
\label{sec:M2}

The second microphone configuration, M2, consists of two concentric circular microphone arrays (CMAs) composed of 4 DPA 4060 and 8 AKG CK 32 microphones. 
Fig. \ref{fig:MicSetupM2} shows a plan view of the M2 configuration, and a description of the microphone labels is given in Table \ref{tab:mic_ls_labels}.
The inner circular microphone array has a radius of \SI{10}{cm} and consists of 4 equidistantly placed DPA 4060 microphones.
The outer circular microphone array has a radius of \SI{20}{cm} and consists of 8 equidistantly placed AKG CK 32 microphones. The microphones are all placed at a height of \SI{1}{\meter} above the floor { using a holder made of laser-cut acrylic glass}, centred around the stand of the DH of the M1 configuration. This M2 configuration   was   used at two different positions within the AIL, always in combination with M1 as depicted in Fig. \ref{fig:mic_ls_placement}. It should be noted that since M2   was   used in combination with M1, it is also possible to define arrays that contain microphones of both configurations, such as a linear array composed of CMA20\_180, CMA10\_180, XM1, CMA10\_0, CMA20\_0, XM2, and XM3.

\subsection{Loudspeaker configurations}
\label{sec:loudspeaker_setups}

\subsubsection{LS-SAL}
\label{sec:LS_SAL}

The loudspeaker configuration LS-SAL as the name suggests is used in the SAL only.
It is defined relative to the M1 microphone configuration, and consists of 10 loudspeakers. 
The loudspeakers are positioned at various spatial locations at a height such that the centre of each of the woofers is approximately \SI{1.3}{\metre} above the floor.
Fig. \ref{fig:MicSetupM1} is a plan view of this LS-SAL loudspeaker configuration along with the M1 microphone configuration. A description of the loudspeaker labels is also provided in Table \ref{tab:mic_ls_labels}. 
During recordings, the loudspeaker S0\_1 was removed before recording the signals for  the loudspeaker S0\_2 so that there   was a direct line of sight from the latter to the DH.

\subsubsection{LS-AIL}
\label{sec:LS_AIL}
 
The loudspeaker configuration LS-AIL is a 24-loud- speaker array, permanently installed in the AIL, cf. Fig. \ref{fig:fisheye_both}, which is typically used for spatial sound reproduction. 
Fig. \ref{fig:speakersAIL} shows the geometry of the loudspeaker array. 
The loudspeakers are labeled as described in Fig. \ref{fig:speakersAIL} and Table \ref{tab:mic_ls_labels}.
The width and depth of the array are approximately \SI{5.6}{\metre} and \SI{4.85}{\metre}, and the loudspeakers are arranged in three groups of different height levels, referred to as lower, upper, and top level.
The lower level consists of 8 speakers located around the room along the walls at about \SI{1.5}{\metre} height, 
the upper level containing 12 speakers is located above at about \SI{3.3}{\metre} height, 
and the top level containing 4 speakers is located more centrally at about \SI{4.1}{\metre} height. 
Note that for the sake of simplicity, the presented locations are only approximate.
Using measurements of the distances between the speakers and a set of four reference points on the floor with known coordinates, the exact coordinates of the loudspeakers   have been   estimated based on the theory on Euclidean distance matrices \cite{Dokmanic15}. 
All microphone and loudspeaker coordinates can be loaded from the database as discussed in Sec. \ref{sec:retrievingcoordinates}.

\subsection{Microphone and loudspeaker configuration placement}
\label{sec:setupplacement}

\begin{figure*}
	\centering
	\includegraphics[scale = 0.365]{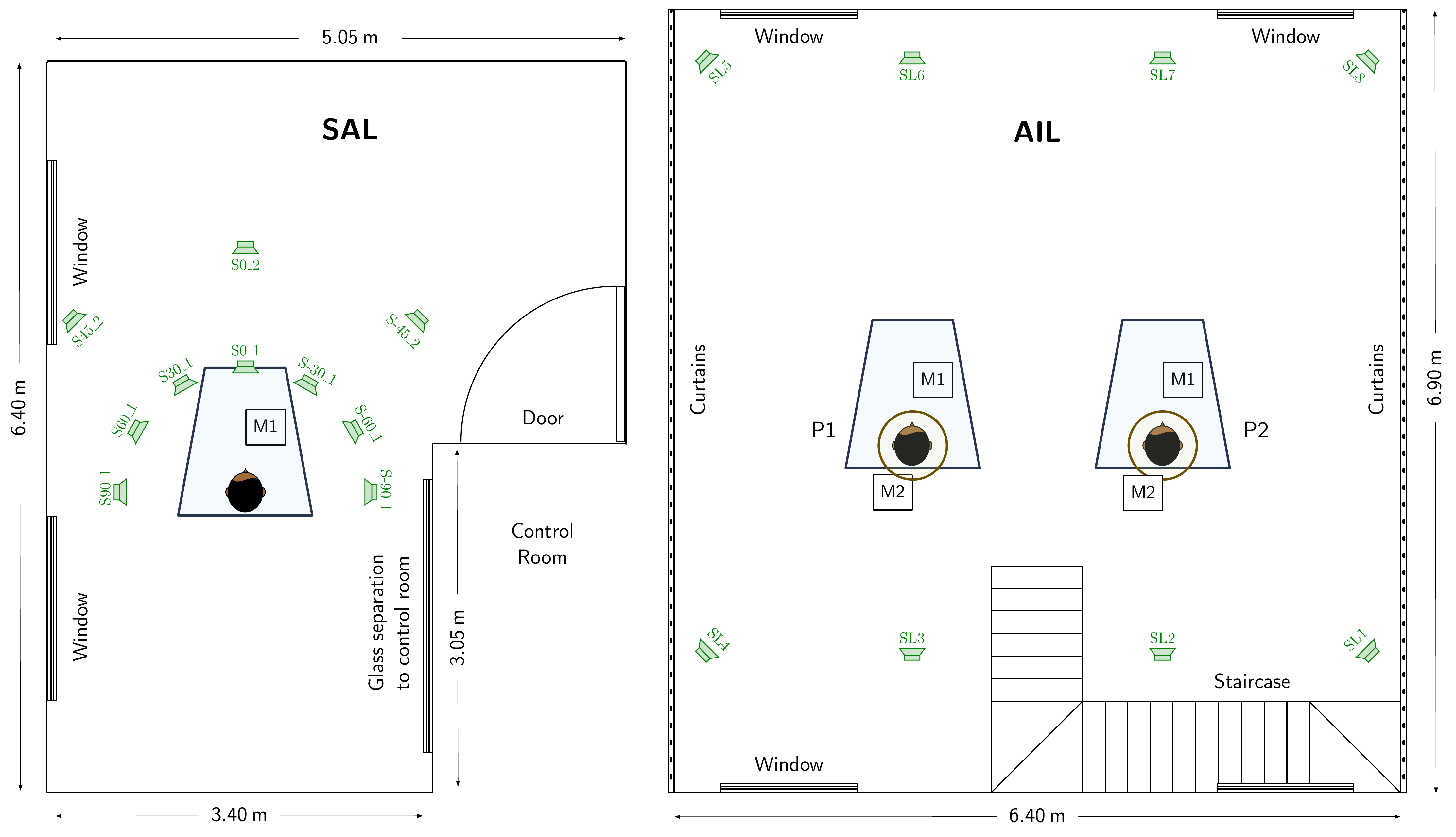}
	\captionsetup{oneside,margin={.16cm, .9cm}}
	\begin{minipage}[t]{2.1\columnwidth}
		\caption{\small
		Microphone and loudspeaker configuration placement. 
(Left) Placement of the M1 microphone configuration and the LS-SAL loudspeaker configuration within the SAL. 
(Right) Placement of the M1 and M2 microphone configurations in P1 and P2 as well as the lower level of the LS-AIL loudspeaker configuration within the AIL. 
		Details of the M1 and M2 microphone configurations and the LS-SAL and LS-AIL loudspeaker configuration can be seen in Fig. \ref{fig:MicSetupM1}, Fig. \ref{fig:MicSetupM2}, and Fig. \ref{fig:speakersAIL}. 
		For extracting the coordinates of the microphone and loudspeaker positions, the MATLAB or Python scripts discussed in Sec. \ref{sec:retrievingcoordinates} should be used.}
	\label{fig:mic_ls_placement}	
	\end{minipage}
\end{figure*}

Fig. \ref{fig:mic_ls_placement} illustrates the placement of the M1 microphone configuration as well as the LS-SAL loudspeaker configuration within the SAL at a recording position near the corner of the L-shaped room. 

\begin{figure}
	\centering
	\includegraphics[scale = 0.355]{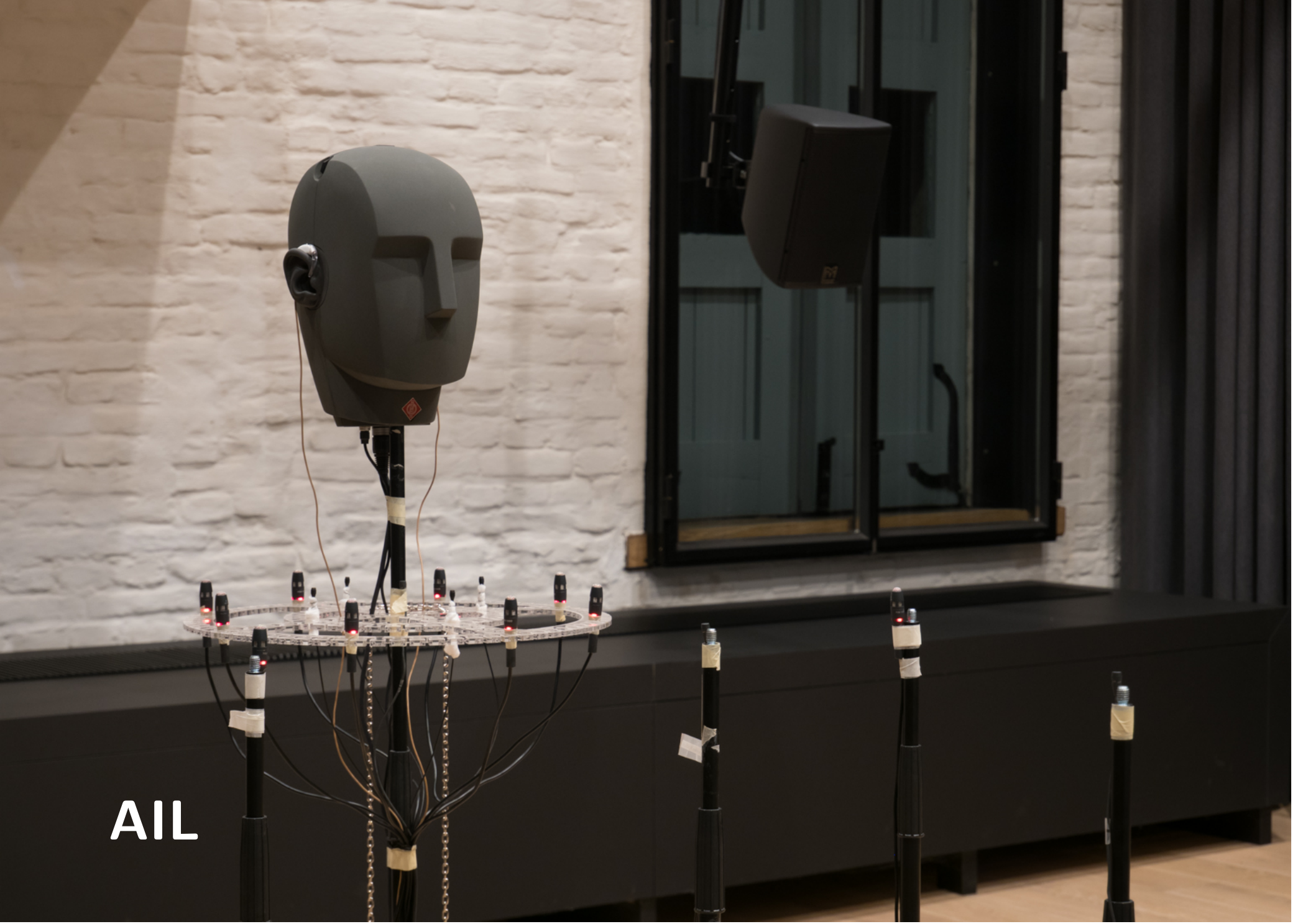}
	\captionsetup{oneside,margin={0.16cm, 0.5cm}}	
	\caption{\small A combination of the microphone configurations M1 and M2 as used at the AIL.}
	\label{fig:mic_config_combination}
\end{figure}

Fig. \ref{fig:mic_ls_placement} shows a floor plan of the setups M1 and M2 within the AIL, together with the lower speakers of the LS-AIL loudspeaker array.
As can be seen, there are two recording positions in the AIL, referred to as P1 and P2, with the DH facing the speakers SU6 and SU7, located roughly below ST2 and ST1 (not shown in the figure), respectively.
In both recording positions, both microphone configurations M1 and M2 are used, with the stand of the DH of M1 being the center of the circular microphone arrays of M2.
Fig. \ref{fig:mic_config_combination} shows a combination of M1 and M2 as used in position P2.

The coordinates of all speakers and microphones in both rooms can be loaded from the database using MATLAB or Python, cf. Sec \ref{sec:retrievingcoordinates}. 

\begin{table*}[t]
	\caption{Signals recorded and computed in the database.}
	\begin{center}
		\label{tab:signals_rec}
		\begin{threeparttable}
			\begin{tabular}{l l l l l l l l l} 
				\toprule
				Signal & Type  & Quantity & Duration (s) & Source & Acquisition  & Speakers$^1$ & Label \\ 
				\midrule
				Male speaker  & Speech & 3 & 30--37 & \cite{VCTK} & playback + record  & L$_\text{sub}$ & M[i],  [i] $\in$ \{1, 2, 3\}\\ 
				Female speaker & Speech & 3 & 30--37 & \cite{VCTK} &   playback + record & L$_\text{sub}$ & F[i],  [i] $\in$ \{1, 2, 3\}\\ 
				Stationary noise & Noise & 1 & 35 &  generated &  playback + record  & L$_\text{sub}$ & SN\\ 
				Cocktail party & Noise & 6 & 600 &  party guests & party + record & none & CP[i],  [i] $\in$ \{1, 2, \dots, 6\}\\ 
				Drums & Music & 1 & 41 & \cite{AECW2011} &   playback + record &  L$_\text{sub}$ & DR \\ 
				Piano & Music & 1 & 35 & \cite{EBU} &  playback + record  &  L$_\text{sub}$ & PI \\ 
				Sine sweep$^2$ & Meas. & 2 & 15 &   generated &  playback + record  & all &  \\
				RIR & RIR & 1 & 2--3 &  sine sweeps & computed  \cite{Holters2009} & all & RIR \\
				\bottomrule
			\end{tabular}
			\begin{tablenotes}
				\item  $^1$ The subset L$_\text{sub}$ includes all speakers in SAL and SL1 to SL8 in the AIL, cf. Fig. \ref{fig:speakersAIL} and Table \ref{tab:mic_ls_labels}. 
				\item $^2$ The raw sine sweeps are not included in the database and hence do not have a label.
			\end{tablenotes}
		\end{threeparttable}
	\end{center}		
\end{table*} 

\begin{figure}[h!]
	\centering
		\includegraphics[scale = 0.355]{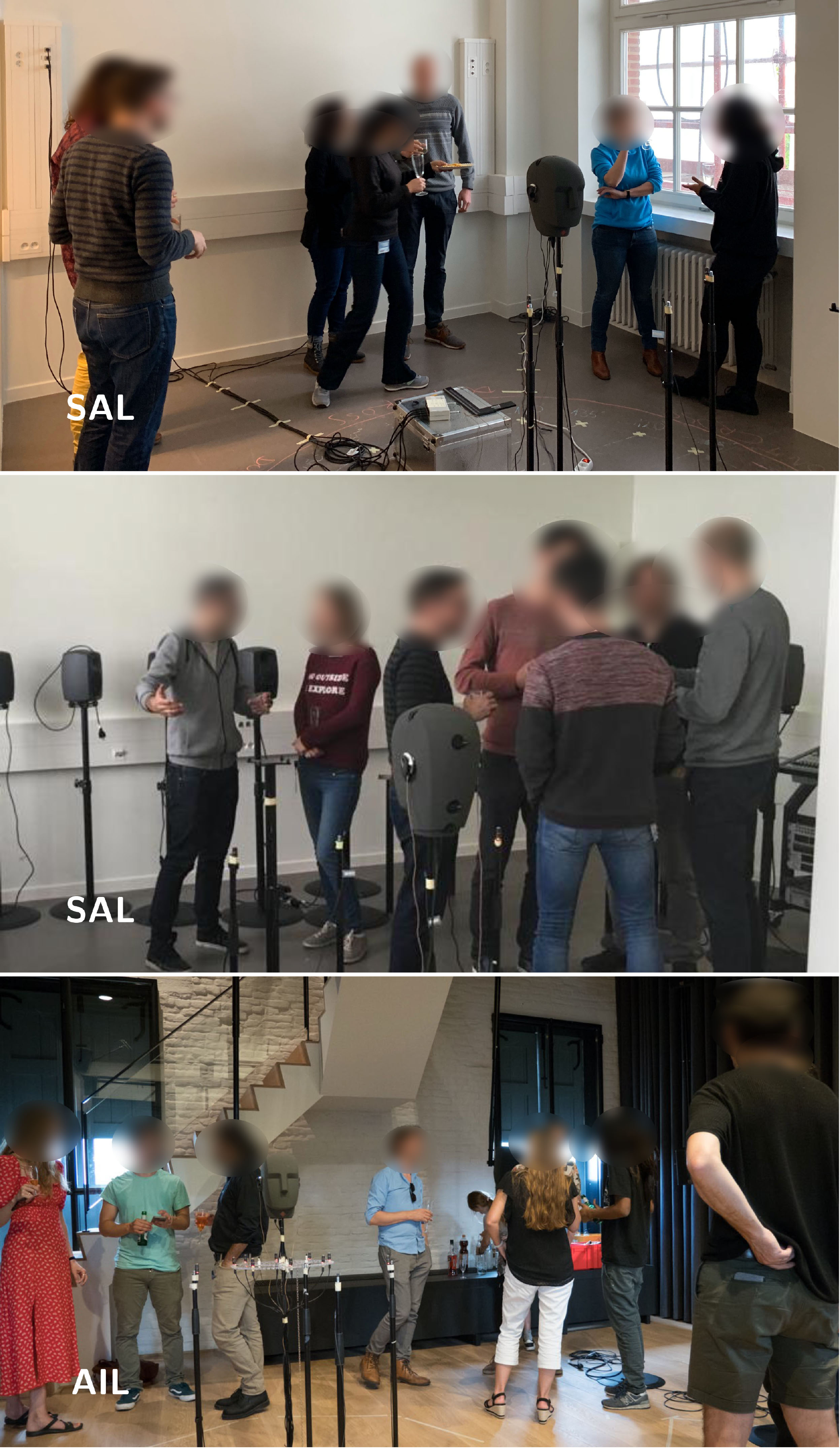}
			\captionsetup{oneside,margin={0.16cm, 0.5cm}}	
	\caption{\small Cocktail party recordings at the SAL and the AIL.}
	\label{fig:cocktailparty_both}
\end{figure}

\section{Recorded signals}
\label{sec:recorded}

The MYRiAD database contains 76 hours of audio data and has a size of \SI{36.2}{GB}.
All microphone signals in the database are provided at a sampling frequency of \SI{44.1}{\kilo\hertz} with a 24 bit resolution. 
{ Their gains are set such that the recording level across the different microphone models is approximately the same around \SI{1}{kHz} in diffuse noise.}
{ For the sake of consistency, recordings were done simultaneously\footnote{This implies acoustic scattering effects that may not match the envisioned application. For instance, when simulations are performed using the CMAs, scattering from the DH may not be meaningful to the simulated scenario. Nevertheless, given that an accurate reproduction of scattering is hardly ever practical, this does not compromise the use of these signals to evaluate acoustic signal processing algorithms.} for all microphones in the SAL as well as in each of the two recording positions P1 and P2 in the AIL.}
A summary of the signals recorded and computed, along with the quantity of each (i.e. the number of different instances of that type of signal), their duration, their source, their acquisition method (i.e. how the signals were generated), the employed loudspeakers, and a signal label is provided in Table \ref{tab:signals_rec}. 
In the remainder of this section, we discuss in more detail the RIR measurements in Sec. \ref{sec:RIR}, the recorded speech, noise and music signals in Sec. \ref{sec:speech_noise_music}, and the recorded cocktail party in Sec. \ref{sec:cocktail_party}.

\subsection{Room impulse responses}
\label{sec:RIR}

The database includes in total 110 RIRs from the SAL and 1104 RIRs from the AIL. 
To obtain the RIRs, 
two exponential sine sweep signals   were played   and recorded for each loudspeaker-microphone combination. In the AIL, the sides of the room were closed off with curtains during the recording.
From these sine sweeps, the RIRs were computed by cross-correlation\footnote{It should be noted that the estimated impulse responses also include some characteristics of the recording hardware. Consequently these impulse responses are, in a strict sense, not the true RIRs which represent the characteristics of the room only. Nevertheless these impulse responses are designated as RIRs for simplicity.} according to the procedure detailed in \cite{Holters2009}. 
From each pair of recorded sine sweeps, one of them was selected for RIR estimation by visual inspection of the spectrograms (more specifically, spectrograms containing any type of non-stationary noise were discarded). 
In order to obtain as clean as possible RIRs, some of the recorded sine sweeps were post-processed as to suppress low-level (stationary) harmonic noise components produced by the recording equipment. 
In this post-processing procedure, frequency bins containing harmonic noise components were identified during silence by comparing their magnitude to the median magnitude of neighbouring frequency bins. If the difference was above the threshold of \SI{4}{dB}, a Wiener filter \cite{loizou2007speech} was applied in that frequency bin. 
The recorded signals were further post-processed to remove the input-output delay caused by the recording hardware.

\subsection{Speech, noise, music}
\label{sec:speech_noise_music} 

Speech, stationary noise, and music signals   were   played through the loudspeakers indicated in Table \ref{tab:signals_rec} and recorded by all microphones.
Three male and three female speech segments were chosen randomly from the Centre for Speech Technology Research (CSTR) Voice Cloning Toolkit (VCTK) corpus \cite{VCTK}. 
The stationary noise source signal has a speech-shaped spectrum and   was   generated in MATLAB based on speech spectra from the VCTK corpus. 
The drum piece  was   taken from the studio recording sessions in \cite{AECW2011}. 
The piano piece is track 60 (Schubert) from the European Broadcast Union Sound Quality Assessment Material Recordings for Subjective Tests (EBU SQAM) \cite{EBU}. In the AIL, the sides of the room were closed off with curtains during recording. 
These signals were acquired for all loudspeakers in the SAL, but only for the lower loudspeaker level in the AIL, that is SL1 to SL8 (in contrast to the RIRs, which were computed for all possible loudspeaker-microphone combinations, cf.  Sec. \ref{sec:RIR}). 
The recorded signals were post-processed to remove the input-output delay caused by the recording hardware.
For the signals recorded in the SAL, a slow phase drift was observed between the recorded data and simulated data obtained from convolving the estimated RIR with the source signal, cf. Sec. \ref{sec:audioexamples}. This phase drift can be associated to hardware limitations in the recording setup and has been compensated for by time-shifting some of the recorded signals\footnote{Only a minority of the recorded signals required a shift of at most 2 samples.} such as to minimize the error between the recorded and the convolved data. 
For the signals recorded in the AIL, no phase drift was observed.
Both the source signals and the recorded signals are included in the database.

\subsection{Cocktail party}
\label{sec:cocktail_party}

In addition to the aforementioned signals, a cocktail party scenario   was   re-created and recorded in both the SAL and the AIL. 
All participants gave informed consent.
They were instructed to stay outside of a \SI{1}{\meter} circumference around the DH in both rooms and periodically move around in a random manner engaging in conversation. Snacks and beverages in glasses were also served to the participants during the recordings. 
For the SAL cocktail party,  at any given time, there were at least 15 people present in the room, whereas for the AIL cocktail party, there were at least 10 and at most 14 people present.
In the SAL, the microphone configuration M1 located as shown in Fig. \ref{fig:mic_ls_placement} was used (the loudspeakers were removed from the room). 
In the AIL, the microphone configurations M1 and M2  located in position P2 as shown in Fig. \ref{fig:mic_ls_placement} were used.
The curtains on the sides of the room in the AIL were closed during the recordings of CP1, CP2, and CP3, and open during CP4, CP5, and CP6.  Photos from the cocktail parties in the SAL and AIL are shown in Fig.~\ref{fig:cocktailparty_both}.

	\begin{table*}[t]
		\caption{File path structure of the database.}
						\begin{center}

		    \begin{threeparttable}
					\begin{tabular}{@{} l l l l l l} 
				\toprule
				& root &  & signal type$^1$\\ \cmidrule{2-4}
				source signal path & /audio/ &SRC/ &[s].wav\\
				\midrule
				\midrule
				& root & room & speaker$^2$ or CP & config. placement$^3$ & microphone$^2$ and signal type$^1$\\ \cmidrule{2-6}
				microphone signal path &/audio/ &SAL/ & S{[a]}\_{[d]}/ & & {[m]}\_{[s]}.wav\\
				\cmidrule{4-4}
			    & & & CP/ & & \\
				\cmidrule{3-5}
				& &AIL/ & S{[l]}{[i]}/ & P1/ & \\
				\cmidrule{5-5}
				& & &  & P2/ & \\
				\cmidrule{4-5}
				& & & CP/ & P2/ & \\
				\midrule
				\midrule
				& root & room &\\ \cmidrule{2-3}
				coordinate file path &coord/ & SAL.csv\\
				 & & AIL.csv\\
				\midrule
				\midrule
				& root & language & script or function$^4$\\ \cmidrule{2-4}
				code file path &/tools/ &MATLAB/ &[f].m\\
				\cmidrule{3-4}
			   & &Python/ & [f].py\\
				\bottomrule
			\end{tabular}
			     \begin{tablenotes}
	           \item $^1$ The signal label [s] takes the forms as defined in Table \ref{tab:signals_rec}.
				\item $^2$  The speaker labels S{[a]}\_{[d]} and S{[l]}{[i]} and the microphone label [m] take the forms as defined in Table \ref{tab:mic_ls_labels}.
				\item $^3$ P1 and P2 refer to the microphone configuration placements at the AIL as shown in Fig. \ref{fig:mic_ls_placement}. 
				\item $^4$ The script or function names [f] take the forms as defined in Table \ref{tab:functions}.
	        	\end{tablenotes}

	     \end{threeparttable}
	        			\end{center}
		\label{tab:file_path}
\end{table*} 

\begin{table*}[t]
	\caption{Scripts facilitating the use of the database.}
	\begin{center}
		\begin{tabular}{@{} l l} 
			\toprule
			script or function name & description (detailed help can be found in the header)\\
			\midrule
			load\_audio\_data &  example script loading audio recordings and calling load\_coordinates()\\
			load\_coordinates() & function loading and optionally plotting microphone and loudspeaker coordinates\\
			\bottomrule
		\end{tabular}
	\end{center}
   \label{tab:functions}
\end{table*}

\begin{figure*}[t]
	\centering
	\includegraphics[scale = 0.79]{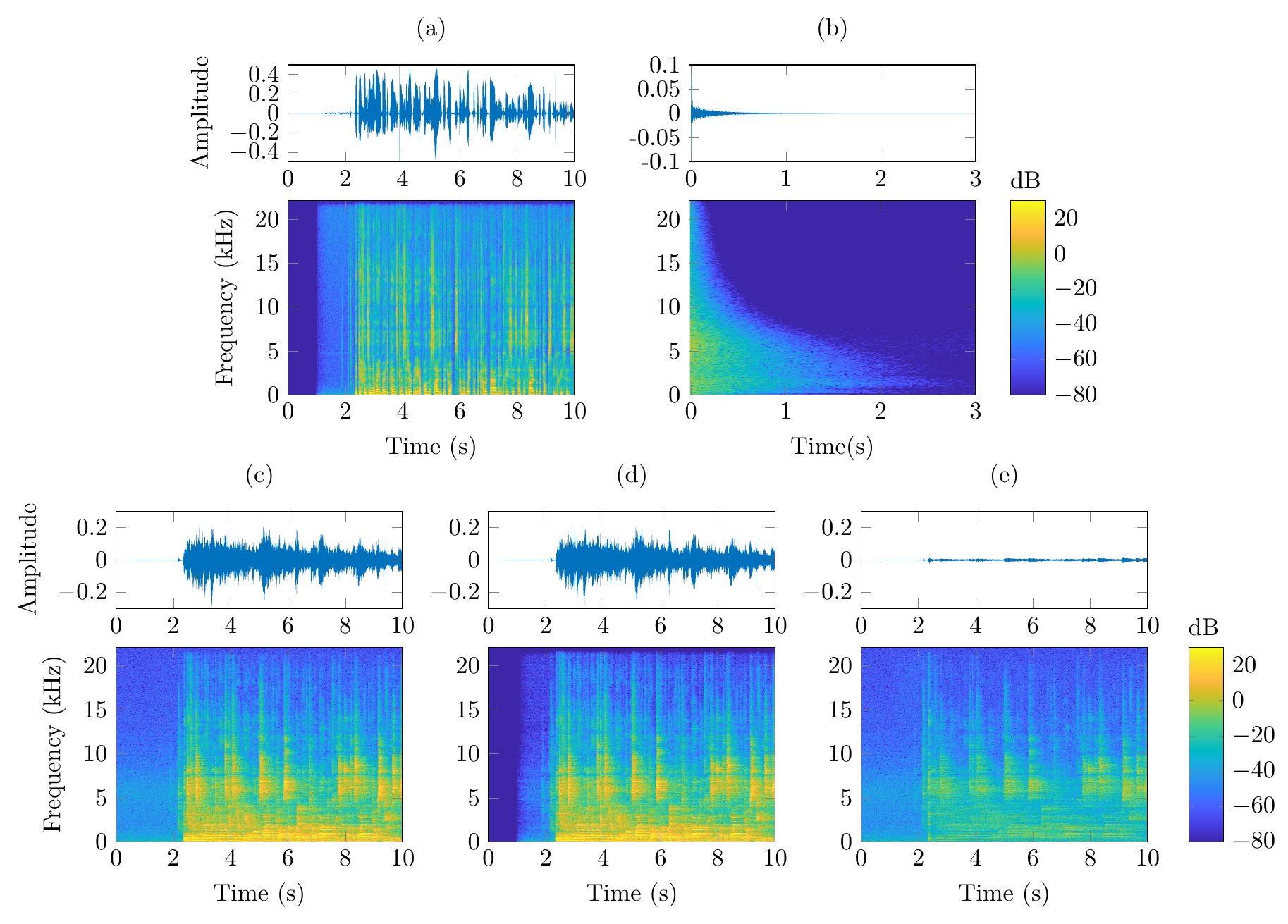}
	\captionsetup{oneside,margin={.16cm, .2cm}}
	\begin{minipage}[t]{2\columnwidth}
		\caption{\small Waveform and corresponding spectrogram of signals related to the SAL recordings. (a) First 10 seconds of the source signal corresponding to a female speaker, F1 (cf. Table \ref{tab:signals_rec}), (b) computed RIR from the loudspeaker S0\_1 to microphone BTELF (cf. Fig. \ref{fig:MicSetupM1}), (c) recorded microphone BTELF signal after the signal from (a) was played through the loudspeaker S0\_1, (d) simulated signal from the convolution of (a) and (b), (e) error between signals (c) and (d).}
			\label{fig:audio_SAL}
	\end{minipage}
\end{figure*}

\begin{figure*}[t]
	\centering
	\includegraphics[scale = 0.79]{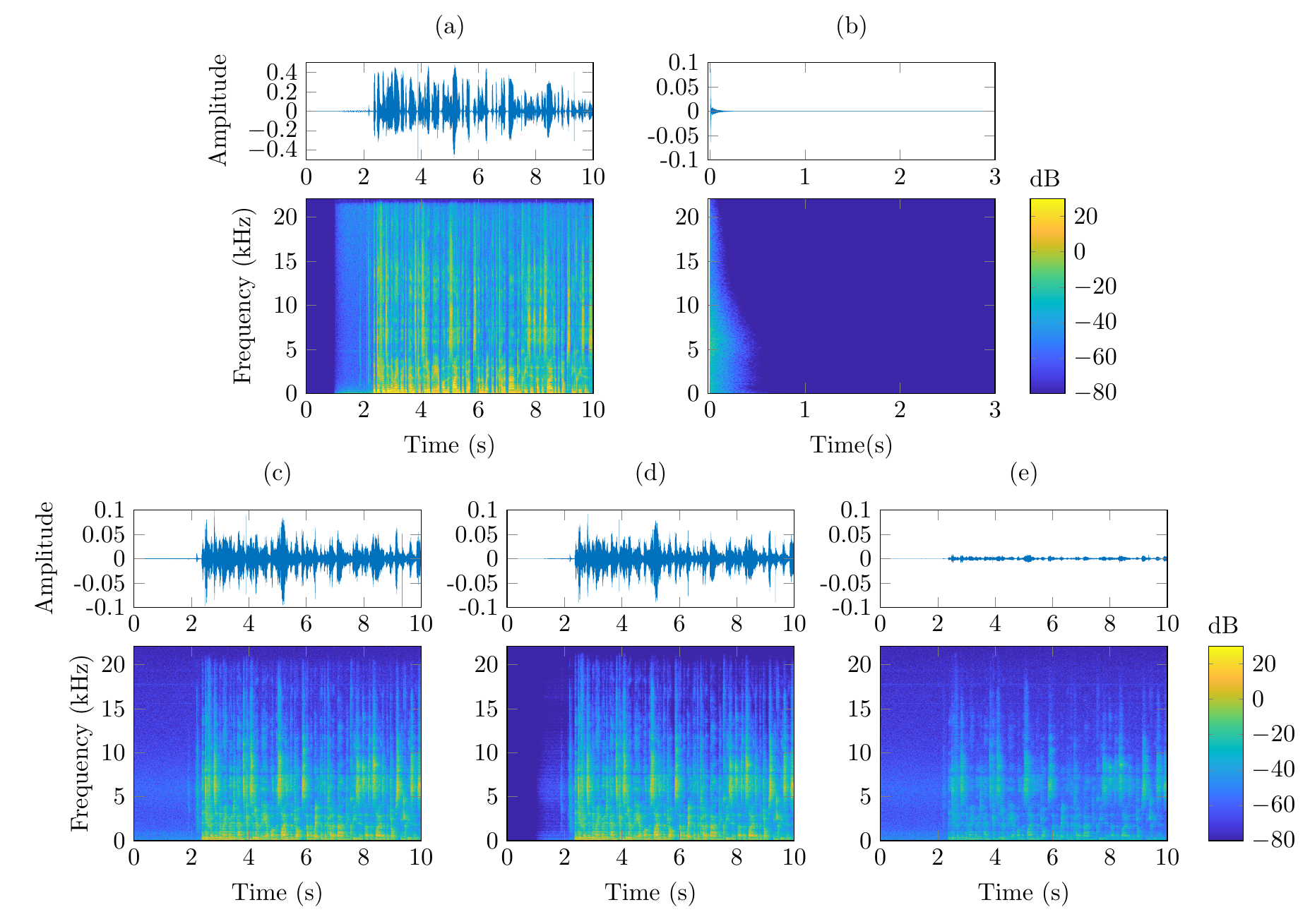}
	\captionsetup{oneside,margin={.16cm, .2cm}}
	\begin{minipage}[t]{2\columnwidth}
		\caption{\small Waveform and corresponding spectrogram of signals related to the AIL recordings. (a) First 10 seconds of the source signal corresponding to a female speaker, F1 (cf. Table \ref{tab:signals_rec}), (b) computed RIR from the loudspeaker SL5\_1 to microphone BTELF (cf. Fig. \ref{fig:MicSetupM1}), (c) recorded microphone BTELF signal after the signal from (a) was played through the loudspeaker SL5\_1, (d) simulated signal from the convolution of (a) and (b), (e) error between signals (c) and (d).}
	\label{fig:audio_AIL}	
	\end{minipage}
\end{figure*}

\begin{figure*}[t]
	\centering
	\includegraphics[scale = 0.79]{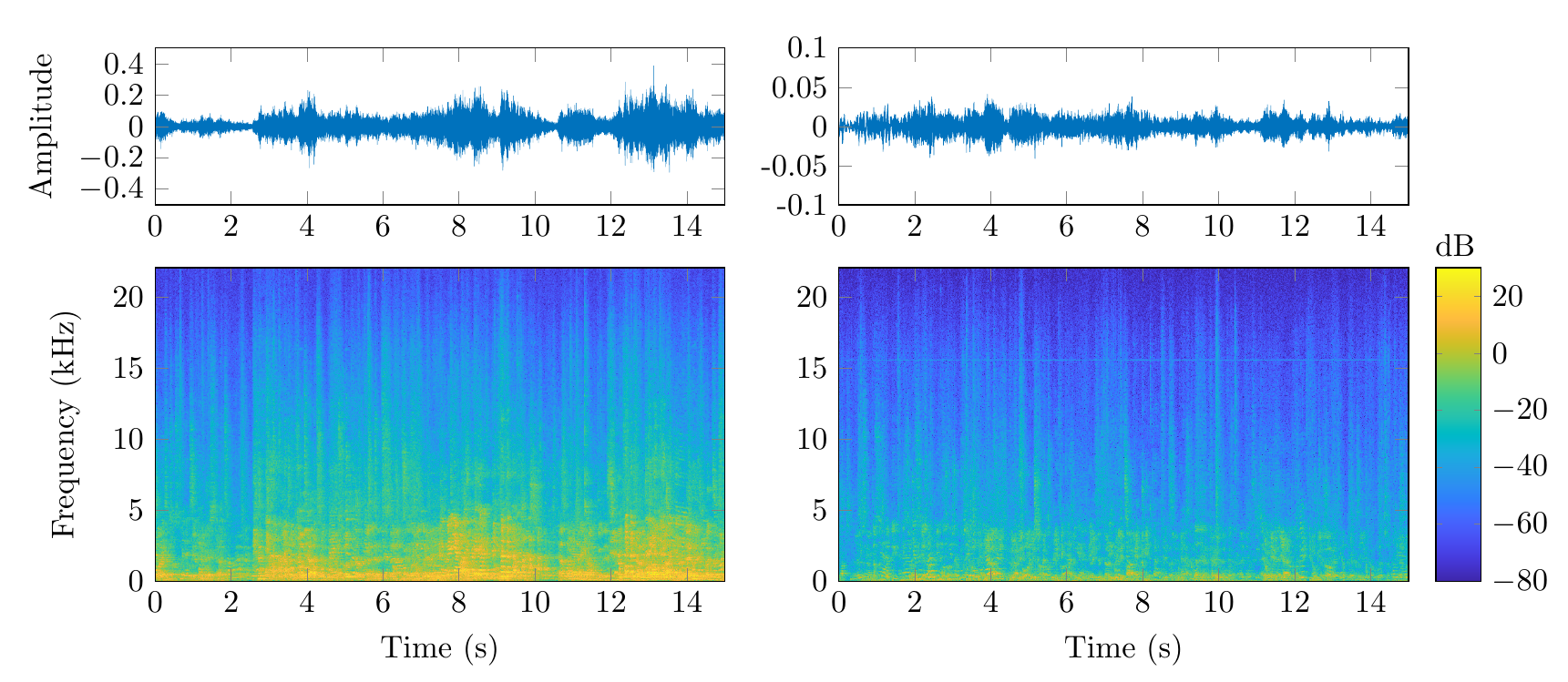}
	\captionsetup{oneside,margin={.16cm, .2cm}}
	\begin{minipage}[t]{2\columnwidth}
		\caption{\small Waveform and corresponding spectrogram for a \SI{15}{\second} sample of the cocktail party noise. (Left) Signal CP2 for XM2 in the SAL. (Right) Signal CP5 for XM2 in the AIL.}
	\label{fig:ckt_noise}	
	\end{minipage}
\end{figure*}

 \begin{figure*}
\centering
\includegraphics[scale = .79]{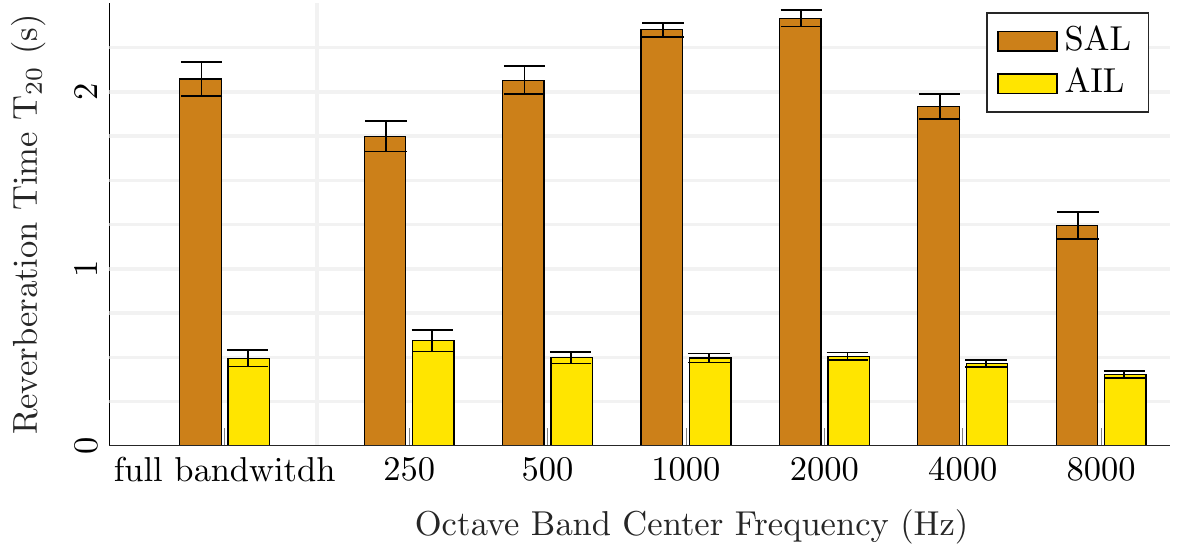}
    \captionsetup{oneside,margin={.16cm, .2cm}}
    \begin{minipage}[t]{2\columnwidth}
		\caption{\small Reverberation time  {$\mathrm{T_{20}}$} for the two rooms SAL and AIL at full bandwidth and in different octave bands. The error bars indicate the standard deviation of the estimate across all possible loudspeaker-microphone combinations.}
		\label{fig:reverbtimes}
	\end{minipage}
\end{figure*}

\section{Using the database}
\label{sec:examples}

In this section, we elaborate on the file path structure of the database in Sec. \ref{sec:filepath} as well as the code provided for loading audio signals and retrieving loudspeaker and microphone coordinates in Sec. \ref{sec:retrievingcoordinates}, and present some examples of audio signals in Sec. \ref{sec:audioexamples} and reverberation time estimates in Sec. \ref{sec:reverbtimes}.

\subsection{File path structure}
\label{sec:filepath}
Table \ref{tab:file_path} provides an overview of the directory tree for the database. Audio files are located in the root directory /audio/, with loudspeaker source signals in the subfolder SRC/ and recorded microphone signals in the subfolders SAL/ and AIL/.
The recorded microphone signals are further organized by loudspeaker (except for cocktail party recordings) and microphone configuration placement (in the AIL). 
The file names encode both the microphone and signal type.
Note that not all folders contain all possible combinations of microphones and signals.
For instance, the folder /audio/SAL/CP/ contains only files of signal type CP$^\ast$, and the folders in /audio/AIL/SU$^\ast$/ and  /audio/AIL/ST$^\ast$/ only contain files of signal type RIR, cf. Sec. \ref{sec:speech_noise_music}.

The folder /coord/ contains files with coordinates of all speakers and microphones in both the SAL and the AIL, and the folder /tools/ contain MATLAB and Python scripts for accessing audio data and coordinates, cf. Sec.\ref{sec:retrievingcoordinates}.

\subsection{Creating Microphone Signals and Retrieving Coordinates}
\label{sec:retrievingcoordinates}

The database comes with MATLAB and Python scripts intended to facilitate retrieving loudspeaker and microphone coordinates and generating signals, as listed in Table \ref{tab:functions}.

The script load\_audio\_data is an example script demonstrating how a .wav-file can be loaded given a list of loudspeaker, microphone, and signal labels provided by the user.
This script also calls the function load\_coordinates(), which reads corresponding coordinates from SAL.csv or AIL.csv (cf. Table \ref{tab:file_path}) and optionally visualizes them.

\subsection{Examples of the audio signals}
\label{sec:audioexamples}

In this section, we take a glimpse into the database by observing some of the signals in both the SAL and the AIL, which will also make evident the different acoustics of the spaces.  

Fig. \ref{fig:audio_SAL} displays the waveform (top of each sub-figure) and corresponding spectrogram (bottom of each sub-figure) for a number of signals related to the SAL. The colourmap in the spectrograms corresponds to the squared magnitude of the short-time Fourier transform coefficients and is plotted in dB. Fig. \ref{fig:audio_SAL} (a) is the first 10 seconds of the source signal corresponding to a female speaker, F1 (cf. Table \ref{tab:signals_rec}). Fig. \ref{fig:audio_SAL} (b) is a computed RIR in the SAL from the loudspeaker S0\_1 to microphone BTELF (cf. Fig. \ref{fig:MicSetupM1}), where the reverberation time is seen to be quite long and highly frequency-dependent. Fig.  \ref{fig:audio_SAL} (c) shows the recorded signal of the source signal F1 (from Fig. \ref{fig:audio_SAL} (a)) in the microphone BTELF after being played through the loudspeaker S0\_1. The effect of the reverberation is evident as the spectrogram shows how the source signal   has now been   distorted in both time and frequency. Fig.  \ref{fig:audio_SAL} (d) is the result of a convolution between the RIR from loudspeaker S0\_1 to microphone BTELF (Fig.  \ref{fig:audio_SAL} (b)) and the F1 source signal (Fig.  \ref{fig:audio_SAL} (a)). This signal is representative of how the recorded signal from  Fig.  \ref{fig:audio_SAL} (c) would typically be simulated. As should be expected, Fig.  \ref{fig:audio_SAL} (c) and Fig.  \ref{fig:audio_SAL} (d), appear quite similar. However, Fig. \ref{fig:audio_SAL} (e) illustrates the difference (error) between the waveform plots in Fig.  \ref{fig:audio_SAL} (c) and Fig.  \ref{fig:audio_SAL} (d), with the corresponding spectrogram of this error, demonstrating that the simulated signal and recorded signal are not identical. 
{ The error may be due to a variety of reasons such as acoustic noise, loudspeaker non-linearities, recording hardware limitations including slow phase drifts, cf. Sec. \ref{sec:speech_noise_music}, and slowly time-variant as well as not perfectly linear sound propagation.}

Fig. \ref{fig:audio_AIL} displays signals from the AIL in a similar manner to that of Fig. \ref{fig:audio_SAL}. The first 10 seconds of the same source signal, F1 (cf. Table \ref{tab:signals_rec}) is observed (Fig. \ref{fig:audio_AIL} (a)).  Fig. \ref{fig:audio_AIL} (b) is a computed RIR in the AIL from the loudspeaker SL5\_1 to microphone BTELF (cf. Fig. \ref{fig:MicSetupM1}), where it can be observed that the reverberation time is significantly shorter as compared to the SAL and more uniform across frequency. Fig.  \ref{fig:audio_AIL} (c) shows the recorded signal of the source signal F1 (from Fig. \ref{fig:audio_AIL} (a)) in the microphone BTELF after being played through the loudspeaker SL5\_1. Fig.  \ref{fig:audio_AIL} (d) is the result of a convolution between the RIR from loudspeaker SL5\_1 to microphone BTELF (Fig.  \ref{fig:audio_AIL} (b)) and the F1 source signal (Fig.  \ref{fig:audio_AIL} (a)).  Fig. \ref{fig:audio_AIL} (e) is the difference (error) between the waveform plots in Fig.  \ref{fig:audio_AIL} (c) and Fig.  \ref{fig:audio_AIL} (d). It can once again be observed that although the simulated and recorded signals are quite similar, they are not identical.

Figure \ref{fig:ckt_noise} depicts the waveform and corresponding spectrogram from a \SI{15}{\second} sample of the cocktail party noise. The left of Fig. \ref{fig:ckt_noise} is the signal CP2 (cf. Table \ref{tab:signals_rec}) for microphone XM2 in the SAL and the right of Fig. \ref{fig:ckt_noise} is the signal CP5 from XM2 in the AIL. The non-stationary behaviour of this type of noise over time and frequency is quite evident.

\subsection{Reverberation times}
\label{sec:reverbtimes}
The reverberation time  {$\mathrm{T_{20}}$} for the two rooms  SAL and AIL is estimated at full bandwidth as well as in different octave bands.
The estimate is obtained from the slope of a line fitted on the decay curves of the RIRs according to the ISO standard \cite{ISO33821} and using the code in \cite{Hummersone2017git}. 
Here, the line   was fitted   in the dynamic range between \SI{-5}{dB} and \SI{-25}{dB} of the decay curve.
A plot of the estimated reverberation times is shown in Fig. \ref{fig:reverbtimes}.
As can be seen, the full-band reverberation time is significantly higher in the SAL with \SI{2.1}{s} as compared to the AIL with \SI{0.5}{s}.
We further note that {$\mathrm{T_{20}}$} in the SAL is largest between \SI{1} and \SI{2}{kHz} and quickly reduces above, while it is less dependent on frequency in the AIL.
{While in the AIL, the variance of the $\mathrm{T_{20}}$ estimates continuously decreases with frequency, we observe that it increases again above to \SI{2}{kHz} in the SAL. 
This may be due to an observed magnitude decay of the SAL RIRs above \SI{2}{kHz}, resulting in less accurate line fitting. In addition, the increased directivity of the loudspeakers at higher frequencies may result in stronger variations of the generated sound field with regards to the loudspeaker placement.}

\section{Conclusion}
\label{sec:conc}

In this paper, a database of acoustic recordings, referred to as the Multi-arraY Room Acoustic Database (MYRiAD), has been presented, which facilitates the recreation of noisy and reverberant microphone signals for the purpose of evaluating audio signal processing algorithms. Recordings were made in two different rooms, the SONORA audio laboratory (SAL) and the Alamire Interactive Laboratory (AIL), with significantly different reverberation times of \SI{2.1}{s} and \SI{0.5}{s}, respectively. In the SAL, a microphone configuration, M1, was used, which consists of in-ear dummy head microphones, microphones on behind-the-ear pieces placed on the dummy head, and external microphones (i.e. other microphones in the room). In the AIL, recordings were made in two different positions within the room using the microphone configuration M1 along with a second microphone configuration, M2, which consists of two concentric circular microphone arrays. In the SAL, 10 movable loudspeakers were used for sound generation, while in the AIL, a built-in array of 24 loudspeakers was used. The database contains room impulse responses, speech, music and stationary noise signals, as well as recordings of a live cocktail party held in each room. MATLAB and Python scripts are included for accessing audio data and coordinates.
The database is publicly available at \cite{Dietzen2022zenodo}.

\section*{Abbreviations}
{MYRiAD --} Multi-ArraY Room Acoustic Database.
{RIR --} room impulse response.
{DH --} dummy head.
{BTE --} behind-the-ear.
{XM --} external microphone.
{CMA --} circular microphone array.
{SAL --} SONORA Audio Laboratory.
{AIL --} Alamire Interactive Laboratory.
{CSTR --} Centre for Speech Technology Research.
{VCTK --} Voice Cloning Toolkit.
{EBU SQAM --} European Broadcast Union Sound Quality Assessment Material.

\section*{Declarations}

\begin{backmatter}

\section*{Availability of data and materials}
The database is publicly available at \cite{Dietzen2022zenodo}.

\section*{Competing interests}
The authors declare that they have no competing interests.

\section*{Funding}

This research work was carried out at the ESAT Laboratory of KU Leuven, in the frame of KU Leuven internal funds C24/16/019 "Distributed Digital Signal Processing for Ad-hoc Wireless Local Area Audio Networking" and VES/19/004, and was funded by the Research Foundation Flanders (FWO-Vlaanderen) through the Large-scale research infrastructure "The Library of Voices -- Unlocking the Alamire Foundation’s Music Heritage Resources Collection through Visual and Sound Technology" (I013218N), the SBO Project "The sound of music -- Innovative research and valorization of plainchant through digital technology" (S005319N), and the Postdoctoral Research Grant 12X6719N. 
The research leading to these results has received funding from the European Research Council under the European Union's Horizon 2020 research and innovation program / ERC Consolidator Grant: SONORA (no. 773268). This paper reflects only the authors' views and the Union is not liable for any use that may be made of the contained information.
The BTE shells used for the recordings were provided by Cochlear Technology Centre Belgium in the context of a project funded by the IWT (project 110722).

\section*{Author's contributions}
TD, RA, MT, and TVW jointly developed the recording setup and methodology.
TD, RA, and MT acquired and post-processed the audio data.
TD and RA compiled the database and drafted the manuscript.
All authors read and reviewed the final manuscript.

\section*{Acknowledgements}
\label{sec:ack}

We would like to thank the European Broadcasting Union for permission to use track 60 (Schubert) from the EBU SQUAM recordings \cite{EBU}.
We would like to thank Rudi Knoops and Jo Santy for coordinating the logistics for recording at the AIL, as well as Elisa Tengan Pires de Souza, Taewoong Lee, and Jesper Brunnstr\"{o}m for assistance with the recordings and preparation of the database. Finally, we would like to thank everyone who participated in the cocktail party noise recordings.

\section*{Authors' information}

MT contributed to this work while being with ESAT-STADIUS, KU Leuven, Leuven,  Belgium.

%\bibliographystyle{bmc-mathphys} 
%\bibliography{audio_lab_refs_withoutDOI}


\newcommand{\BMCxmlcomment}[1]{}

\BMCxmlcomment{

<refgrp>

<bibl id="B1">
  <title><p>Data Repository for {MYRiAD}: A Multi-Array Room Acoustic
  Database</p></title>
  <aug>
    <au><snm>Dietzen</snm><fnm>T.</fnm></au>
    <au><snm>Ali</snm><fnm>R.</fnm></au>
    <au><snm>Taseska</snm><fnm>M.</fnm></au>
    <au><snm>Waterschoot</snm><fnm>T.</fnm></au>
  </aug>
  <publisher>Zenodo</publisher>
  <pubdate>2022</pubdate>
  <url>https://zenodo.org/record/7389996</url>
</bibl>

<bibl id="B2">
  <title><p>Speech enhancement: theory and practice</p></title>
  <aug>
    <au><snm>Loizou</snm><fnm>P. C.</fnm></au>
  </aug>
  <publisher>Boca Raton, Florida, USA: CRC press</publisher>
  <pubdate>2007</pubdate>
</bibl>

<bibl id="B3">
  <title><p>Springer Handbook of Speech Processing</p></title>
  <aug>
    <au><snm>Gannot</snm><fnm>S.</fnm></au>
    <au><snm>Cohen</snm><fnm>I.</fnm></au>
  </aug>
  <publisher>New York City, New York State, USA: Springer</publisher>
  <section><title><p>Adaptive Beamforming and
  Postfiltering</p></title></section>
  <pubdate>2007</pubdate>
  <fpage>945</fpage>
  <lpage>-978</lpage>
</bibl>

<bibl id="B4">
  <title><p>Acoustic Beamforming for Hearing Aid Applications</p></title>
  <aug>
    <au><snm>Doclo</snm><fnm>S.</fnm></au>
    <au><snm>Gannot</snm><fnm>S.</fnm></au>
    <au><snm>Moonen</snm><fnm>M.</fnm></au>
    <au><snm>Spriet</snm><fnm>A.</fnm></au>
  </aug>
  <source>Handbook on Array Processing and Sensor Networks</source>
  <publisher>Hoboken, New Jersey, USA: Wiley</publisher>
  <pubdate>2010</pubdate>
  <fpage>269</fpage>
  <lpage>-302</lpage>
</bibl>

<bibl id="B5">
  <title><p>Speech Dereverberation</p></title>
  <aug>
    <au><snm>Naylor</snm><fnm>P. A.</fnm></au>
    <au><snm>Gaubitch</snm><fnm>N. D.</fnm></au>
  </aug>
  <publisher>New York City, New York State, USA: Springer</publisher>
  <pubdate>2010</pubdate>
</bibl>

<bibl id="B6">
  <title><p>Microphone arrays: signal processing techniques and
  applications</p></title>
  <aug>
    <au><snm>Brandstein</snm><fnm>M.</fnm></au>
    <au><snm>Ward</snm><fnm>D.</fnm></au>
  </aug>
  <publisher>New York City, New York State, USA: Springer</publisher>
  <pubdate>2013</pubdate>
</bibl>

<bibl id="B7">
  <title><p>A summary of the {REVERB} challenge: state-of-the-art and remaining
  challenges in reverberant speech processing research</p></title>
  <aug>
    <au><snm>Kinoshita</snm><fnm>K.</fnm></au>
    <au><snm>Delcroix</snm><fnm>M.</fnm></au>
    <au><snm>Gannot</snm><fnm>S.</fnm></au>
    <au><snm>Habets</snm><fnm>E. A. P.</fnm></au>
    <au><snm>Haeb Umbach</snm><fnm>R.</fnm></au>
    <au><snm>Kellermann</snm><fnm>W.</fnm></au>
    <au><snm>Leutnant</snm><fnm>V.</fnm></au>
    <au><snm>Maas</snm><fnm>R.</fnm></au>
    <au><snm>Nakatani</snm><fnm>T.</fnm></au>
    <au><snm>Raj</snm><fnm>B.</fnm></au>
    <au><snm>Sehr</snm><fnm>A.</fnm></au>
    <au><snm>Yoshioka</snm><fnm>T.</fnm></au>
  </aug>
  <source>EURASIP J. Adv. Signal Process.</source>
  <publisher>Springer</publisher>
  <pubdate>2016</pubdate>
  <volume>2016</volume>
  <issue>7</issue>
  <fpage>1</fpage>
  <lpage>-19</lpage>
</bibl>

<bibl id="B8">
  <title><p>A consolidated perspective on multimicrophone speech enhancement
  and source separation</p></title>
  <aug>
    <au><snm>Gannot</snm><fnm>S.</fnm></au>
    <au><snm>Vincent</snm><fnm>E.</fnm></au>
    <au><snm>Markovich Golan</snm><fnm>S.</fnm></au>
    <au><snm>Ozerov</snm><fnm>A.</fnm></au>
  </aug>
  <source>IEEE/ACM Trans. Audio, Speech, Lang. Process.</source>
  <publisher>IEEE</publisher>
  <pubdate>2017</pubdate>
  <volume>25</volume>
  <issue>4</issue>
  <fpage>692</fpage>
  <lpage>-730</lpage>
</bibl>

<bibl id="B9">
  <title><p>Audio Source Separation and Speech Enhancement</p></title>
  <aug>
    <au><snm>Vincent</snm><fnm>E.</fnm></au>
    <au><snm>Virtanen</snm><fnm>T.</fnm></au>
    <au><snm>Gannot</snm><fnm>S.</fnm></au>
  </aug>
  <publisher>Hoboken, New Jersey, USA: Wiley</publisher>
  <pubdate>2018</pubdate>
</bibl>

<bibl id="B10">
  <title><p>Estimation of room acoustic parameters: The {ACE}
  Challenge</p></title>
  <aug>
    <au><snm>Eaton</snm><fnm>J.</fnm></au>
    <au><snm>Gaubitch</snm><fnm>N. D.</fnm></au>
    <au><snm>Moore</snm><fnm>A. H.</fnm></au>
    <au><snm>Naylor</snm><fnm>P. A.</fnm></au>
  </aug>
  <source>IEEE/ACM Trans. Audio, Speech, Lang. Process.</source>
  <publisher>IEEE</publisher>
  <pubdate>2016</pubdate>
  <volume>24</volume>
  <issue>10</issue>
  <fpage>1681</fpage>
  <lpage>-1693</lpage>
</bibl>

<bibl id="B11">
  <title><p>{ICASSP 2021} acoustic echo cancellation challenge: Datasets,
  testing framework, and results</p></title>
  <aug>
    <au><snm>Sridhar</snm><fnm>K.</fnm></au>
    <au><snm>Cutler</snm><fnm>R.</fnm></au>
    <au><snm>Saabas</snm><fnm>A.</fnm></au>
    <au><snm>Parnamaa</snm><fnm>T.</fnm></au>
    <au><snm>Loide</snm><fnm>M.</fnm></au>
    <au><snm>Gamper</snm><fnm>H.</fnm></au>
    <au><snm>Braun</snm><fnm>S.</fnm></au>
    <au><snm>Aichner</snm><fnm>R.</fnm></au>
    <au><snm>Srinivasan</snm><fnm>S.</fnm></au>
  </aug>
  <source>Proc. 2021 IEEE Int. Conf. Acoust., Speech, Signal Process. (ICASSP
  2021)</source>
  <publisher>Toronto, Ontario, Canada</publisher>
  <pubdate>2021</pubdate>
  <fpage>151</fpage>
  <lpage>-155</lpage>
</bibl>

<bibl id="B12">
  <title><p>Fifty years of acoustic feedback control: State of the art and
  future challenges</p></title>
  <aug>
    <au><snm>Waterschoot</snm><fnm>T.</fnm></au>
    <au><snm>Moonen</snm><fnm>M.</fnm></au>
  </aug>
  <source>Proc. IEEE</source>
  <publisher>IEEE</publisher>
  <pubdate>2010</pubdate>
  <volume>99</volume>
  <issue>2</issue>
  <fpage>288</fpage>
  <lpage>-327</lpage>
</bibl>

<bibl id="B13">
  <title><p>The {LOCATA} challenge: Acoustic source localization and
  tracking</p></title>
  <aug>
    <au><snm>Evers</snm><fnm>C.</fnm></au>
    <au><snm>L{\"o}llmann</snm><fnm>H. W.</fnm></au>
    <au><snm>Mellmann</snm><fnm>H.</fnm></au>
    <au><snm>Schmidt</snm><fnm>A.</fnm></au>
    <au><snm>Barfuss</snm><fnm>H.</fnm></au>
    <au><snm>Naylor</snm><fnm>P. A.</fnm></au>
    <au><snm>Kellermann</snm><fnm>W.</fnm></au>
  </aug>
  <source>IEEE/ACM Trans. Audio, Speech, Lang. Process.</source>
  <publisher>IEEE</publisher>
  <pubdate>2020</pubdate>
  <volume>28</volume>
  <fpage>1620</fpage>
  <lpage>-1643</lpage>
</bibl>

<bibl id="B14">
  <title><p>Acoustic contrast, planarity and robustness of sound zone methods
  using a circular loudspeaker array</p></title>
  <aug>
    <au><snm>Coleman</snm><fnm>P.</fnm></au>
    <au><snm>Jackson</snm><fnm>P. J. B.</fnm></au>
    <au><snm>Olik</snm><fnm>M.</fnm></au>
    <au><snm>M{\o}ller</snm><fnm>M.</fnm></au>
    <au><snm>Olsen</snm><fnm>M.</fnm></au>
    <au><snm>Abildgaard Pedersen</snm><fnm>J.</fnm></au>
  </aug>
  <source>J. Acoust. Soc. Am.</source>
  <publisher>Acoustical Society of America</publisher>
  <pubdate>2014</pubdate>
  <volume>135</volume>
  <issue>4</issue>
  <fpage>1929</fpage>
  <lpage>-1940</lpage>
</bibl>

<bibl id="B15">
  <title><p>Personal sound zones: Delivering interface-free audio to multiple
  listeners</p></title>
  <aug>
    <au><snm>Betlehem</snm><fnm>T.</fnm></au>
    <au><snm>Zhang</snm><fnm>W.</fnm></au>
    <au><snm>Poletti</snm><fnm>M. A.</fnm></au>
    <au><snm>A.</snm><fnm>T. D.</fnm></au>
  </aug>
  <source>IEEE Signal Process. Mag.</source>
  <publisher>IEEE</publisher>
  <pubdate>2015</pubdate>
  <volume>32</volume>
  <issue>2</issue>
  <fpage>81</fpage>
  <lpage>-91</lpage>
</bibl>

<bibl id="B16">
  <title><p>The Fifth '{CHiME}' Speech Separation and Recognition Challenge:
  Dataset, Task and Baselines</p></title>
  <aug>
    <au><snm>Barker</snm><fnm>J.</fnm></au>
    <au><snm>Watanabe</snm><fnm>S.</fnm></au>
    <au><snm>Vincent</snm><fnm>E.</fnm></au>
    <au><snm>Trmal</snm><fnm>J.</fnm></au>
  </aug>
  <source>Proc. Interspeech 2018</source>
  <publisher>Hyderabad, India</publisher>
  <pubdate>2018</pubdate>
  <fpage>1561</fpage>
  <lpage>-1565</lpage>
</bibl>

<bibl id="B17">
  <title><p>Evaluation of speech dereverberation algorithms using the {MARDY}
  database</p></title>
  <aug>
    <au><snm>Wen</snm><fnm>J. Y. C.</fnm></au>
    <au><snm>Gaubitch</snm><fnm>N. D.</fnm></au>
    <au><snm>Habets</snm><fnm>E. A. P.</fnm></au>
    <au><snm>Myatt</snm><fnm>T.</fnm></au>
    <au><snm>Naylor</snm><fnm>P. A.</fnm></au>
  </aug>
  <source>Proc. 2006 Intl. Workshop Acoust. Echo Noise Control (IWAENC
  2006)</source>
  <publisher>Paris, France</publisher>
  <pubdate>2006</pubdate>
</bibl>

<bibl id="B18">
  <title><p>A binaural room impulse response database for the evaluation of
  dereverberation algorithms</p></title>
  <aug>
    <au><snm>Jeub</snm><fnm>M.</fnm></au>
    <au><snm>Schafer</snm><fnm>M.</fnm></au>
    <au><snm>Vary</snm><fnm>P.</fnm></au>
  </aug>
  <source>Proc. 2009 16th Int. Conf. Digital Signal Process. (DSP
  2009)</source>
  <publisher>Santorini, Greece</publisher>
  <pubdate>2009</pubdate>
  <fpage>1</fpage>
  <lpage>5</lpage>
</bibl>

<bibl id="B19">
  <title><p>{Database of multichannel in-ear and behind-the-ear head-related
  and binaural room impulse responses}</p></title>
  <aug>
    <au><snm>Kayser</snm><fnm>H.</fnm></au>
    <au><snm>Ewert</snm><fnm>S. D.</fnm></au>
    <au><snm>Anem{\"{u}}ller</snm><fnm>J.</fnm></au>
    <au><snm>Rohdenburg</snm><fnm>T.</fnm></au>
    <au><snm>Hohmann</snm><fnm>V.</fnm></au>
    <au><snm>Kollmeier</snm><fnm>B.</fnm></au>
  </aug>
  <source>EURASIP J. Adv. Signal Process.</source>
  <pubdate>2009</pubdate>
  <volume>2009</volume>
</bibl>

<bibl id="B20">
  <title><p>Database of omnidirectional and {B}-format room impulse
  responses</p></title>
  <aug>
    <au><snm>Stewart</snm><fnm>R.</fnm></au>
    <au><snm>Sandler</snm><fnm>M.</fnm></au>
  </aug>
  <source>Proc. 2010 IEEE Int. Conf. Acoust., Speech, Signal Process. (ICASSP
  2010)</source>
  <publisher>Dallas, Texas, USA</publisher>
  <pubdate>2010</pubdate>
  <fpage>165</fpage>
  <lpage>-168</lpage>
</bibl>

<bibl id="B21">
  <title><p>The single- and multichannel audio recordings database
  ({SMARD})</p></title>
  <aug>
    <au><snm>Nielsen</snm><fnm>J. K.</fnm></au>
    <au><snm>Jensen</snm><fnm>J. R.</fnm></au>
    <au><snm>Jensen</snm><fnm>S. H.</fnm></au>
    <au><snm>Christensen</snm><fnm>M. G.</fnm></au>
  </aug>
  <source>Proc. 2014 Int. Workshop Acoustic Signal Enhancement (IWAENC
  2014)</source>
  <publisher>Antibes -- Juan les Pins, France</publisher>
  <pubdate>2014</pubdate>
  <fpage>40</fpage>
  <lpage>-44</lpage>
</bibl>

<bibl id="B22">
  <title><p>Multichannel audio database in various acoustic
  environments</p></title>
  <aug>
    <au><snm>Hadad</snm><fnm>E.</fnm></au>
    <au><snm>Heese</snm><fnm>F.</fnm></au>
    <au><snm>Vary</snm><fnm>P.</fnm></au>
    <au><snm>Gannot</snm><fnm>S.</fnm></au>
  </aug>
  <source>Proc. 2014 Int. Workshop Acoustic Signal Enhancement (IWAENC
  2014)</source>
  <publisher>Antibes -- Juan les Pins, France</publisher>
  <pubdate>2014</pubdate>
  <fpage>313</fpage>
  <lpage>-317</lpage>
</bibl>

<bibl id="B23">
  <title><p>A real-world recording database for ad hoc microphone
  arrays</p></title>
  <aug>
    <au><snm>Woods</snm><fnm>WS</fnm></au>
    <au><snm>Hadad</snm><fnm>E</fnm></au>
    <au><snm>Merks</snm><fnm>I</fnm></au>
    <au><snm>Xu</snm><fnm>B</fnm></au>
    <au><snm>Gannot</snm><fnm>S</fnm></au>
    <au><snm>Zhang</snm><fnm>T</fnm></au>
  </aug>
  <source>Proc. 2015 IEEE Workshop Appl. Signal Process. Audio, Acoust. (WASPAA
  2015)</source>
  <publisher>New Paltz, NY, USA</publisher>
  <pubdate>2015</pubdate>
  <fpage>2</fpage>
  <lpage>-6</lpage>
</bibl>

<bibl id="B24">
  <title><p>Building and evaluation of a real room impulse response
  dataset</p></title>
  <aug>
    <au><snm>Sz{\"o}ke</snm><fnm>I.</fnm></au>
    <au><snm>Sk{\'a}cel</snm><fnm>M.</fnm></au>
    <au><snm>Mo{\v{s}}ner</snm><fnm>L.</fnm></au>
    <au><snm>Paliesek</snm><fnm>J.</fnm></au>
    <au><snm>{\v{C}}ernock{\`y}</snm><fnm>J.</fnm></au>
  </aug>
  <source>IEEE J. Selected Topics Signal Process.</source>
  <publisher>IEEE</publisher>
  <pubdate>2019</pubdate>
  <volume>13</volume>
  <issue>4</issue>
  <fpage>863</fpage>
  <lpage>-876</lpage>
</bibl>

<bibl id="B25">
  <title><p>d{E}chorate: a calibrated room impulse response dataset for
  echo-aware signal processing</p></title>
  <aug>
    <au><snm>Di Carlo</snm><fnm>D.</fnm></au>
    <au><snm>Tandeitnik</snm><fnm>P.</fnm></au>
    <au><snm>Foy</snm><fnm>C.</fnm></au>
    <au><snm>Bertin</snm><fnm>N.</fnm></au>
    <au><snm>Deleforge</snm><fnm>A.</fnm></au>
    <au><snm>Gannot</snm><fnm>S.</fnm></au>
  </aug>
  <source>EURASIP J. Audio, Speech, Music Process.</source>
  <publisher>Springer</publisher>
  <pubdate>2021</pubdate>
  <volume>2021</volume>
  <issue>1</issue>
  <fpage>1</fpage>
  <lpage>-15</lpage>
</bibl>

<bibl id="B26">
  <title><p>{MIR}a{G}e: multichannel database of room impulse responses
  measured on high-resolution cube-shaped grid</p></title>
  <aug>
    <au><snm>{\v{C}}mejla</snm><fnm>J.</fnm></au>
    <au><snm>Kounovsk{\`y}</snm><fnm>T.</fnm></au>
    <au><snm>Gannot</snm><fnm>S.</fnm></au>
    <au><snm>Koldovsk{\`y}</snm><fnm>Z.</fnm></au>
    <au><snm>Tandeitnik</snm><fnm>P.</fnm></au>
  </aug>
  <source>2020 28th European Signal Process. Conf. (EUSIPCO 2020)</source>
  <publisher>Amsterdam, The Netherlands</publisher>
  <pubdate>2021</pubdate>
  <fpage>56</fpage>
  <lpage>-60</lpage>
</bibl>

<bibl id="B27">
  <title><p>{MESHRIR}: A Dataset of Room Impulse Responses on Meshed Grid
  Points for Evaluating Sound Field Analysis and Synthesis Methods</p></title>
  <aug>
    <au><snm>Koyama</snm><fnm>S.</fnm></au>
    <au><snm>Nishida</snm><fnm>T.</fnm></au>
    <au><snm>Kimura</snm><fnm>K.</fnm></au>
    <au><snm>Abe</snm><fnm>T.</fnm></au>
    <au><snm>Ueno</snm><fnm>N.</fnm></au>
    <au><snm>Brunnstr{\"o}m</snm><fnm>J.</fnm></au>
  </aug>
  <source>Proc. 2021 IEEE Workshop Appl. Signal Process. Audio, Acoust. (WASPAA
  2021)</source>
  <publisher>New Paltz, NY, USA: {IEEE}</publisher>
  <pubdate>2021</pubdate>
  <fpage>1</fpage>
  <lpage>-5</lpage>
</bibl>

<bibl id="B28">
  <title><p>A room impulse response database for multizone sound field
  reproduction</p></title>
  <aug>
    <au><snm>Zhao</snm><fnm>S.</fnm></au>
    <au><snm>Zhu</snm><fnm>Q.</fnm></au>
    <au><snm>Cheng</snm><fnm>E.</fnm></au>
    <au><snm>Burnett</snm><fnm>I. S.</fnm></au>
  </aug>
  <source>J. Acoust. Soc. Am.</source>
  <publisher>Acoustical Society of America</publisher>
  <pubdate>2022</pubdate>
  <volume>152</volume>
  <issue>4</issue>
  <fpage>2505</fpage>
  <lpage>-2512</lpage>
</bibl>

<bibl id="B29">
  <title><p>{D}i{PC}o - Dinner Party Corpus</p></title>
  <aug>
    <au><snm>Van Segbroeck</snm><fnm>M.</fnm></au>
    <au><snm>Zaid</snm><fnm>A.</fnm></au>
    <au><snm>Kutsenko</snm><fnm>K.</fnm></au>
    <au><snm>Huerta</snm><fnm>C.</fnm></au>
    <au><snm>Nguyen</snm><fnm>T.</fnm></au>
    <au><snm>Luo</snm><fnm>X.</fnm></au>
    <au><snm>Hoffmeister</snm><fnm>B.</fnm></au>
    <au><snm>Trmal</snm><fnm>J.</fnm></au>
    <au><snm>Omologo</snm><fnm>M.</fnm></au>
    <au><snm>Maas</snm><fnm>R.</fnm></au>
  </aug>
  <source>arXiv preprint arXiv:1909.13447</source>
  <pubdate>2019</pubdate>
</bibl>

<bibl id="B30">
  <title><p>{Multichannel acoustic source and image dataset for the cocktail
  party effect in hearing aid and implant users}</p></title>
  <aug>
    <au><snm>Fischer</snm><fnm>T.</fnm></au>
    <au><snm>Caversaccio</snm><fnm>M.</fnm></au>
    <au><snm>Wimmer</snm><fnm>W.</fnm></au>
  </aug>
  <source>Scientific Data</source>
  <pubdate>2020</pubdate>
  <volume>7</volume>
  <issue>440</issue>
  <fpage>1</fpage>
  <lpage>-13</lpage>
</bibl>

<bibl id="B31">
  <title><p>Informed Sound Source Localization Using Relative Transfer
  Functions for Hearing Aid Applications</p></title>
  <aug>
    <au><snm>Farmani</snm><fnm>M.</fnm></au>
    <au><snm>Pedersen</snm><fnm>M. S.</fnm></au>
    <au><snm>Tan</snm><fnm>Z. H.</fnm></au>
    <au><snm>Jensen</snm><fnm>J.</fnm></au>
  </aug>
  <source>IEEE/ACM Trans. Audio, Speech, Language Process.</source>
  <pubdate>2017</pubdate>
  <volume>25</volume>
  <issue>3</issue>
  <fpage>611</fpage>
  <lpage>623</lpage>
</bibl>

<bibl id="B32">
  <title><p>Performance Analysis of the Extended Binaural {MVDR} Beamformer
  With Partial Noise Estimation</p></title>
  <aug>
    <au><snm>G\"o{\ss}ling</snm><fnm>N.</fnm></au>
    <au><snm>Marquardt</snm><fnm>D.</fnm></au>
    <au><snm>Doclo</snm><fnm>S.</fnm></au>
  </aug>
  <source>IEEE/ACM Trans. Audio, Speech, Language Process.</source>
  <pubdate>2021</pubdate>
  <volume>29</volume>
  <fpage>462</fpage>
  <lpage>-476</lpage>
</bibl>

<bibl id="B33">
  <title><p>Multi-microphone speech enhancement : An integration of a priori
  and data-dependent spatial information</p></title>
  <aug>
    <au><snm>Ali</snm><fnm>R.</fnm></au>
  </aug>
  <source>PhD thesis</source>
  <publisher>KU Leuven, Leuven, Belgium</publisher>
  <pubdate>2020</pubdate>
</bibl>

<bibl id="B34">
  <title><p>On the Design of Frequency-Invariant Beampatterns With Uniform
  Circular Microphone Arrays</p></title>
  <aug>
    <au><snm>Huang</snm><fnm>G.</fnm></au>
    <au><snm>Benesty</snm><fnm>J.</fnm></au>
    <au><snm>Chen</snm><fnm>J.</fnm></au>
  </aug>
  <source>IEEE/ACM Trans. Audio, Speech, Language Process.</source>
  <pubdate>2017</pubdate>
  <volume>25</volume>
  <issue>5</issue>
  <fpage>1140</fpage>
  <lpage>-1153</lpage>
</bibl>

<bibl id="B35">
  <title><p>Real-Time Multiple Sound Source Localization and Counting Using a
  Circular Microphone Array</p></title>
  <aug>
    <au><snm>Pavlidi</snm><fnm>D.</fnm></au>
    <au><snm>Griffin</snm><fnm>A.</fnm></au>
    <au><snm>Puigt</snm><fnm>M.</fnm></au>
    <au><snm>Mouchtaris</snm><fnm>A.</fnm></au>
  </aug>
  <source>IEEE Trans. Audio, Speech, Language Process.</source>
  <pubdate>2013</pubdate>
  <volume>21</volume>
  <issue>10</issue>
  <fpage>2193</fpage>
  <lpage>-2206</lpage>
</bibl>

<bibl id="B36">
  <title><p>KU Leuven ESAT-STADIUS Audio Research Labs</p></title>
  <aug>
    <au><snm>Waterschoot</snm><fnm>T.</fnm></au>
  </aug>
  <source>\url{https://lirias.kuleuven.be/3940173}</source>
  <pubdate>2022</pubdate>
</bibl>

<bibl id="B37">
  <title><p>{Impulse response measurement techniques and their applicability in
  the real world}</p></title>
  <aug>
    <au><snm>Holters</snm><fnm>M.</fnm></au>
    <au><snm>Corbach</snm><fnm>T.</fnm></au>
    <au><snm>Z{\"{o}}lzer</snm><fnm>U.</fnm></au>
  </aug>
  <source>Proc. 2009 12th Int. Conf. Digital Audio Effects (DAFx 2009)</source>
  <publisher>Como, Italy</publisher>
  <pubdate>2009</pubdate>
  <fpage>108</fpage>
  <lpage>-112</lpage>
</bibl>

<bibl id="B38">
  <title><p>{CSTR} {VCTK} corpus: English multi-speaker corpus for {CSTR} voice
  cloning toolkit</p></title>
  <aug>
    <au><snm>Veaux</snm><fnm>C.</fnm></au>
    <au><snm>Yamagishi</snm><fnm>J.</fnm></au>
    <au><snm>MacDonald</snm><fnm>K.</fnm></au>
  </aug>
  <source>\url{http://homepages.
  inf.ed.ac.uk/jyamagis/page3/page58/page58.html}</source>
  <pubdate>2016</pubdate>
</bibl>

<bibl id="B39">
  <title><p>Federation {D}ay</p></title>
  <aug>
    <au><cnm>Anti Everything</cnm></au>
  </aug>
  <source>Children of a Globalised World (Musical Album)</source>
  <note>{ISRC}: TTA101100005, Boatshrimp Records, Port-of-Spain, 2011</note>
</bibl>

<bibl id="B40">
  <title><p>Sound Quality Assessment Material Recordings for Subjective
  Tests</p></title>
  <aug>
    <au><snm>Union</snm><fnm>EB</fnm></au>
  </aug>
  <source>\url{https://tech.ebu.ch/publications/sqamcd}</source>
  <pubdate>2008</pubdate>
</bibl>

<bibl id="B41">
  <title><p>Euclidean Distance Matrices: Essential Theory, Algorithms and
  Applications</p></title>
  <aug>
    <au><snm>Dokmani{\'c}</snm><fnm>I.</fnm></au>
    <au><snm>Parhizkar</snm><fnm>R.</fnm></au>
    <au><snm>Ranieri</snm><fnm>J.</fnm></au>
    <au><snm>Vetterli</snm><fnm>M.</fnm></au>
  </aug>
  <source>IEEE Signal Process. Mag.</source>
  <pubdate>2015</pubdate>
  <volume>32</volume>
  <fpage>12</fpage>
  <lpage>30</lpage>
</bibl>

<bibl id="B42">
  <title><p>Acoustics — Measurement of room acoustic parameters — Part 1:
  Performance spaces</p></title>
  <aug>
    <au><cnm>ISO{~}3382 1:2009</cnm></au>
  </aug>
  <publisher>Geneva, Switzerland: International Organization for
  Standardization</publisher>
  <pubdate>2009</pubdate>
  <fpage>26</fpage>
</bibl>

<bibl id="B43">
  <title><p>Git{H}ub Repository: {I}o{SR} {Matlab} Toolbox</p></title>
  <aug>
    <au><snm>Hummersone</snm><fnm>C.</fnm></au>
    <au><snm>Pr\"{a}tzlich</snm><fnm>T.</fnm></au>
  </aug>
  <source>\url{https://github.com/IoSR-Surrey/MatlabToolbox}</source>
  <pubdate>2017</pubdate>
</bibl>

</refgrp>
} % end of \BMCxmlcomment

\end{backmatter}
\end{document}